\documentclass[abstract]{aa}

\usepackage{graphicx}
\usepackage[dvipsnames]{xcolor}
\usepackage{txfonts}
\usepackage{xspace}
\usepackage{float}
\usepackage{url}
\usepackage{multirow}
\usepackage{gensymb}
\graphicspath{{images/}}
\usepackage{makecell}
\usepackage{appendix}
\usepackage{lipsum}  
\usepackage{upgreek}
\usepackage[colorlinks=True, citecolor=Blue, urlcolor=blue, pagebackref=True, breaklinks=false]{hyperref}
\usepackage{orcidlink}
\renewcommand*{\backref}[1]{}
\renewcommand*{\backrefalt}[4]{[{\tiny%
    \ifcase #1 Not cited.%
          \or Cited on page~#2%
          \else Cited on pages #2%
    \fi%
    }].}

\newcommand{\g}{$\upgamma$\xspace}
\newcommand{\co}{$\rm{CO}$\xspace}
\newcommand{\twco}{$^{12}\rm{CO}$\xspace}
\newcommand{\twcoi}{$^{12}\rm{CO}(1-0)$\xspace}
\newcommand{\twcoii}{$^{12}\rm{CO}(2-1)$\xspace}
\newcommand{\thco}{$^{13}\rm{CO}$\xspace}
\newcommand{\thcoi}{$^{13}\rm{CO}(1-0)$\xspace}
\newcommand{\thcoii}{$^{13}\rm{CO}(2-1)$\xspace}
\newcommand{\eico}{$\rm{C}^{18}\rm{O}$\xspace}
\newcommand{\eicoii}{$\rm{C}^{18}\rm{O}(2-1)$\xspace}
\newcommand{\wco}{$W_{\rm{CO}}$\xspace}
\newcommand{\wtwco}{$W_{^{12}\rm{CO}}$\xspace}
\newcommand{\wtwcoi}{$W_{^{12}\rm{CO}}^{10}$\xspace}
\newcommand{\wtwcoii}{$W_{^{12}\rm{CO}}^{21}$\xspace}

\newcommand{\wthcoi}{$W_{^{13}\rm{CO}}^{10}$\xspace}
\newcommand{\wthcoii}{$W_{^{13}\rm{CO}}^{21}$\xspace}
\newcommand{\weico}{$W_{\rm{C}^{18}\rm{O}}$\xspace}
\newcommand{\weicoii}{$W_{\rm{C}^{18}\rm{O}}^{21}$\xspace}
\newcommand{\wcounit}{K km s$^{-1}$\xspace}

\newcommand{\hd}{H$_2$\xspace}

\newcommand{\hi}{{\sc Hi}\xspace}

\newcommand{\mh}{$m_{\rm{H}}$\xspace}
\newcommand{\nh}{$N_{\rm{H}}$\xspace}
\newcommand{\nhi}{$N_{\rm \sc HI}$\xspace}

\newcommand{\nhd}{$N_{\rm{H}_2}$\xspace}

\newcommand{\opa}{$\tau_{353}/N_{\rm{H}}$\xspace}

\newcommand{\opaunit}{$10^{-26}$ cm$^2$\xspace}

\newcommand{\xco}{$X_{\rm{CO}}$\xspace}
\newcommand{\xcotwenty}{$X_{\rm{CO}, 20}$\xspace}

\newcommand{\xcounit}{$10^{20}$ cm$^{-2}$ K$^{-1}$ km$^{-1}$ s\xspace}
\newcommand{\taunu}{$\tau_{353}$\xspace}

\newcommand{\chisq}{$\chi^2 $\xspace}

\newcommand{\tauratio}{$\tau_{\rm 353}^{\rm cold}/\tau_{\rm 353}^{\rm warm}$\xspace}

\begin{document}

   \title{Cosmic rays, gas and dust in the Central Molecular Zone \\
   I -- \xco factors, cosmic-ray densities and dust opacities} 

\author{H.~X.~Ren$^{(1)}$ \orcidlink{0000-0003-0221-2560} \and 
Q.~Remy$^{(1)}$ \orcidlink{0000-0002-8815-6530} \and 
S.~Ravikularaman$^{(2)}$ \orcidlink{0000-0002-2974-1668} \and 
M.~Bouyahiaoui$^{(1)}$ \and
F.~Conte$^{(1)}$ \orcidlink{0000-0002-3083-8539} \and
J.~Djuvsland$^{(1, 3)}$ \orcidlink{0000-0002-6488-8219} 
 }

\institute{
~Max-Planck-Institut für Kernphysik, P.O. Box 103980, D 69029 Heidelberg, Germany \and
~Université de Paris Cité, CNRS, Astroparticule et Cosmologie, F-75013 Paris, France \and
~Department for Physics and Technology, University of Bergen, Allegaten 55, Bergen, 5020, Norway\\
\email{helena.ren@mpi-hd.mpg.de}\\
\email{quentin.remy@mpi-hd.mpg.de}
}

\date{Received 4 February 2025 / Accepted 14 March 2025}

\abstract
{The Central Molecular Zone (CMZ) is a unique environment in our Galaxy, with extreme conditions to test our understanding of the gas, dust and cosmic-ray (CR) physics. 
}
{Our goal is to estimate the total gas mass in the direction of the Galactic centre (GC), quantify the various uncertainties associated, and discuss the implications for the estimates of CR energy densities and dust opacities.} 
{The \hi 21 cm line and the carbon monoxide isotopes (\twco ($J = 1 \rightarrow 0$) , \thco ($J = 1 \rightarrow 0$; $J = 2 \rightarrow 1$) and \eico ($J = 2 \rightarrow 1$)) line emission maps are used to derive the total gas column density.
The gas in the CMZ is separated from the disk contribution in position and velocity thanks to its different properties in term of velocity dispersion and brightness ratio of \co isotopes. The variations of the \xco factors are modelled relying on both theoretical trends from simulations and empirical corrections. We use the new gas column density estimated together with gamma-ray and dust emission measurements to derive the CR energy density and dust opacities, respectively.}
{The \xco values in the CMZ range from \mbox{$(0.32$ -- $1.37) \ \times$ \xcounit}, with a distribution that is highly asymmetric and skewed. The median value is \mbox{$ \rm{\overline{X}_{CO}^{CMZ}} = 0.39 \ \times$  \xcounit}.
The total gas mass in the CMZ is estimated to be \mbox{$2.3_{-0.3}^{+0.3}\times10^{7} \; \rm{M_{\odot}}$} with $\sim 10 \%$ contribution from the atomic phase. Without removing the disk contamination the total mass is about twice higher, and the atomic gas fraction increases to $\sim30\%$. 
The CR energy density in the CMZ, assuming a 1/r profile, is higher by a factor of two compared to the previous calculations at TeV energies.
}
{Toward the GC the contamination from both atomic and molecular gas in the disk is not negligible. Using molecular gas tracers which probes only the densest molecular cores leads to an overestimation of the CR energy density,  while ignoring the foreground/background contribution leads to an underestimation of the CR energy density in the CMZ.
}

\titlerunning{Gas \& dust in the CMZ}
\authorrunning{Ren et al.}

\maketitle

\renewcommand{\arraystretch}{2.}
\setlength{\tabcolsep}{0.2cm}

\begin{table*}[ht]
    \centering
    \caption{Overview of the data used in this study.}
    \resizebox{\textwidth}{!}{
    \begin{tabular}{c|c|c|c|c|c|c|c}
        \hline \hline
        \thead{Tracer} & \thead{Frequency} & \thead{$\mathbf{v_{\rm LSR}}$ range} & \thead{$\mathbf{v_{\rm LSR}}$ resolution: \\ effective  (channel)} & \thead{Angular resolution: \\ beam size (pixel) }& \thead{Sensitivity \\ (rms noise)} & \thead{Instrument} & \thead{Reference}\\

        \ & $[\mathrm{GHz}]$ & $[\mathrm{km}\,\mathrm{s}^{-1}]$ &  $[\mathrm{km}\,\mathrm{s}^{-1}]$ & $[\degree]$ & $[\mathrm{K}]$ &  & \\ 
                
        \hline
        {\sc Hi}\xspace & $1.4$ & $-309:349$ & 1  (0.82) & 0.0403  (0.0097)  &   0.7 -- 2  &ATCA & \cite{2012ApJS..199...12M} \\
        \hline
        $^{12}\rm CO (J=1-0)$ & $115.27$ & $-220:220$ & 1.3  (2) &  0.0043  (0.0021) & 1 &45m NRO-BEARS& \cite{2019PASJ...71S..19T} \\
        \hline
        $^{13}\rm CO (J=1-0)$ & $110.20$  & $-220:220$ & 1.3  (2) &  0.0043  (0.0021) & 0.2 &45m NRO-FOREST & \cite{2019PASJ...71S..19T} \\  
        \hline
        $^{13}\rm CO (J=2-1)$ & $220.40$  & $-200:200$ & 0.25  (0.25) &  0.0083  (0.0026) & 0.8 -- 1 &APEX & \cite{2021MNRAS.500.3064S} \\ 
        \hline
        $\rm C^{18}\rm O (J=2-1)$ & $219.56$  & $-200:200$ & 0.25  (0.25) & 0.0083  (0.0026) & 0.8 -- 1 &APEX & \cite{2021MNRAS.500.3064S} \\ 
        \hline  \hline

        \thead{Tracer} & \thead{Wavelength} &  &  & \thead{Angular resolution \\ beam size (pixel)}& \thead{Sensitivity \\ (rms noise)} & \thead{Instrument} & \thead{Reference}\\ 

        \ & $[\mathrm{\upmu m}]$ &   &   & $[\degree]$ & $[\rm{mJy/beam}]$ &  & \\ 
        
        \hline
        Dust emission &$70, 160$ & & & 0.0023, 0.0033  $(\rm{beam}/2.66)$ & Fig 3 of Ref. &Herschel-PACS &  \cite{2016AA...591A.149M} \\ 
        \hline
        Dust emission &$250, 350, 500$ & & & 0.005, 0.0067, 0.0096  $(\rm{beam}/3)$ &  Fig 3 of Ref.  &Herschel-SPIRE & \cite{2016AA...591A.149M} \\ 
        \hline
        Dust emission &$870$ & & & 0.0053  (0.0017) & 70 -- 90 &APEX + \textit{Planck} & \cite{2016AA...585A.104C} \\
        \hline
         Dust emission &$350, 550, 850, 1382, 2096, 2998$ & & & 0.070--0.162 $(0.029)$ & Table 12 of Ref.  & \textit{Planck}-HFI & \cite{2020AandA...641A...3P} \\ 
        \hline \hline

    \end{tabular}}
    \label{tab:tracers}
\end{table*}

\section{Introduction}

The Central Molecular Zone (CMZ) in the Galactic centre (GC) is an extreme environment, unlike any other region in the Galactic disk \citep[see Table 2 of][and references therein for an overview]{hen23}.
It extends up to $\sim$250 pc in positive and $\sim$150 pc in negative longitudes \citep{fer07}. 
This region concentrates a very large fraction of dense molecular gas ($> 10^3 \rm cm^{-3}$), the kinematic temperature of the gas is 5 to 10 times higher than in the disk \citep{1985A&A...142..381G,  gin16}, the clouds have high turbulent velocities \citep{bal87, 1996ARA&A..34..645M}, and the majority of stars have super-solar metallicity \citep{2015ApJ...809..143D,  2018MNRAS.478.4374N, 2019A&A...627A.152S}. Characterising the properties of the various gas tracers in this environment is important to estimate the quantity of gas precisely.

Although \hd is the most abundant molecule in the Universe, it is very difficult to detect \citep{wak17, 2022PhT....75l..12S}. Therefore, many surveys of the molecular gas in the CMZ use other molecules as tracers of \hd. These include the different line transitions of \co isotopologues, which probe most of the gas, HCN or CS for the densest cores, and many of other rarer molecular species \citep[see][and references therein]{2017arXiv170505332M}. However, integrating observations from different molecular tracers, each with varying sensitivities, coverage, and angular resolution, into a single coherent model of the gas in the CMZ is challenging.

The estimate of gas column densities and mass is affected by a number of uncertainties and unknowns. Regarding the molecular phase, several questions can be asked. Up to which density range does the \co emission trace the \hd gas reliably? How does the \co-to-\hd conversion factor, \xco, vary within a cloud and from cloud-to-cloud at parsec scales \citep{2013MNRAS.431.1296R, 2017A&A...601A..78R, 2024PASJ...76..579K}? How does \xco vary across galaxies from their centres \citep{2020A&A...635A.131I, 2023ApJ...950..119T} to kilo-parsec scales \citep{1996PASJ...48..275A, 2013ApJ...777....5S}? How can we model these variations depending on the properties of the environment and the resolved scale \citep{2006MNRAS.371.1865B, 2011MNRAS.412..337G, 2016MNRAS.455.3763B, 2020ApJ...903..142G}? What information can be gained considering multiple molecular gas tracers? How to combine them in order to probe a wider range of densities, on the one hand in the densest core where \co emission saturates \citep{1998A&A...331..959D}, and on the other hand at the transition between atomic and molecular phases where \co surveys lack sensitivity to detect diffuse \hd \citep{2005Sci...307.1292G, 2023A&A...675A.145L}? 
Regarding the atomic phase traced by the neutral hydrogen, \hi, the main sources of uncertainties are the impact of the \hi self-absorption \citep{2022ApJ...929..136P} and the value of the \hi spin temperature \citep{2017MNRAS.468.4030S}.

If one tries to estimate the quantity of gas from observations of dust emission, other questions arise: how well is the dust mixed with the gas? On which spatial scales does the dust-to-gas mass ratio vary and by how much \citep{2015A&A...578A.131L, 2017MNRAS.471L..52T}? How does the dust temperature vary along the line of sight? How do the dust emission properties vary with dust grain evolution in different environments, and how does this affect the estimate of the dust mass \citep{2014A&A...571A..11P, 2015A&A...579A..15K, 2019A&A...631A..88Y, 2024A&A...684A..34Y}? 

Rather than detailing all these questions here, we are going to start by extracting one conclusion: in order to estimate properly the gas column densities, not only does one have to rely on multiple independent tracers, taking into account the variations of their various properties and conversion factors with the environment, but also to take into account the variations with the sensitivity and resolved scale of the observations. As this is not straightforward to apply in data analysis, simpler approaches relying on single tracers and generic constant conversion factors are still used, especially outside the interstellar medium community, like in very-high-energy \g-ray astronomy.

For example, \citet{2018A&A...612A...9H} shows that the \g-ray emission seen by the High Energy Stereoscopic System (H.E.S.S.) in the CMZ is correlated with dense gas traced by CS molecules and that its flux is about half the total diffuse emission flux in the region. Still, the large-scale diffuse component was derived from an empirical model and not based on a gas template.
Earlier studies by \citet{1998A&A...331..959D} showed that the diffuse gas component traced by \twco, outside of the dense cores where \eico is detected, accounts to up to half of the total mass. Therefore, the more diffuse regions can contain as much gas as the denser and cooler cloud cores. Then, the large scale diffuse emission seen in \g-ray by H.E.S.S. toward the GC could also arise from hadronic collisions of cosmic rays (CR) in less dense regions of the interstellar medium (ISM). 
Another study of the GC by \citet{2016Natur.531..476H} considered \co, CS, and HCN as alternative tracers. Still, they have not attempted to combine them to probe a larger range of gas density, and cloud-to-cloud variations of the conversion factors to molecular hydrogen have not been considered so far. Even in the recent study by \cite{2020A&A...642A.190M} only CS emission was considered as gas tracer.
Still nowadays, the lack of a reliable total gas column density model and a clear view of the uncertainties associated is a limiting factor for the \g-ray studies on the CR density profile toward the GC, the base of the Fermi-bubbles, or a potential dark matter excess.

In this paper, we produce a map of the total hydrogen column density, \nh, in the CMZ combining \hi data and a set of \co isotopologue observations. We also aim to quantify the various uncertainties inherent to the exercise. The final products are, in particular, meant to be used in future \g-ray studies of the GC.
The data used are listed in Sect. \ref{sec:data}. The method is summarised as follows: the \hi and \co velocity profiles are decomposed into a collection of lines (Sect. \ref{sec:decomposition}); the properties of the lines are used to separate the CMZ from the gas in the disk (Sect. \ref{sec:disk_sep}); relying on both the theoretical trend from simulations and the empirical corrections we derive \xco factors for each line such as we minimize the variance of \nhd obtained from the different \co isotopologues considered (Sect. \ref{sec:nhd}).
The results on the properties of \co lines are presented in Sect. \ref{sec:res_co}, which includes the discussion about the variations of \xco factors (Sect. \ref{sec:xco}).
The hydrogen column density maps, separated into its different phases (atomic and molecular) and components (CMZ and disk), are shown in Sect. \ref{sec:res_maps}, and associated mass estimates are discussed in Sect. \ref{sec:mass}.
In Sect. \ref{sec:dust_discussion}, the total hydrogen column density map is compared to the distribution of the dust optical depth to check if the variation of the dust opacity is consistent with dust model expectations and comparable to observed trends seen in other parts of the Galaxy.
In Sect. \ref{sec:CR}, we discuss the impact of our new gas mass estimates on the CR densities derived in the GC with respect to previous measurements from the H.E.S.S. Collaboration.
Finally, we summarize the results in Sect. \ref{sec:ccl} and draw some perspectives for the follow-up studies in preparation or future works.

\section{Data} \label{sec:data}

The details of the observations we used are given in Table \ref{tab:tracers}. 
Further details on each individual tracer are given in the following sections.
The analysed region covers longitudes $-0.8\degree< l <1.4\degree$ and latitudes $|b|<0.3\degree$, the region is limited by the coverage of the high resolution \co survey used. 

Part of the data processing is done exploiting the original map resolutions, but in the end, all the maps are downsampled to a common angular resolution of $0.02\degree$. We chose this bin size because it corresponds to a resolved scale slightly larger than the minimum value probed by the simulations we are using to model the \xco evolution. As the final maps are meant to be used in the context of \g-ray astronomy, this resolution is sufficient. Indeed, current \g-ray instruments have coarser resolutions, and only future instruments could reach this level of precision.

\subsection{Gas tracers}

The \hi 21 cm line traces the gas in atomic neutral phase.
We use the \hi Galactic Center survey data conducted with the Australia Telescope Compact Array (ATCA) and the Parkes Radio Telescope \citep{2012ApJS..199...12M}.

The molecular phase is commonly traced by the carbon monoxide emission lines. Three different isotopes (\twco, \thco and \eico) and two rotational transitions ($J = 1 \rightarrow 0$ and $J = 2 \rightarrow 1$, hereafter noted $(1-0)$ and $(2-1)$, respectively).
Based on the survey coverage, resolution and sensitivity, we choose the \twcoi and \thcoi lines from the survey of the CMZ obtained using the 45m telescope at the Nobeyama Radio Observatory (NRO) \citep{2019PASJ...71S..19T}, and the \thcoii and \eicoii lines from the SEDIGISM survey conducted with the Atacama Pathfinder Experiment (APEX) 12m submillimetre telescope \citep{2021MNRAS.500.3064S}.

\begin{figure}
    \centering
    \includegraphics[width=\columnwidth]{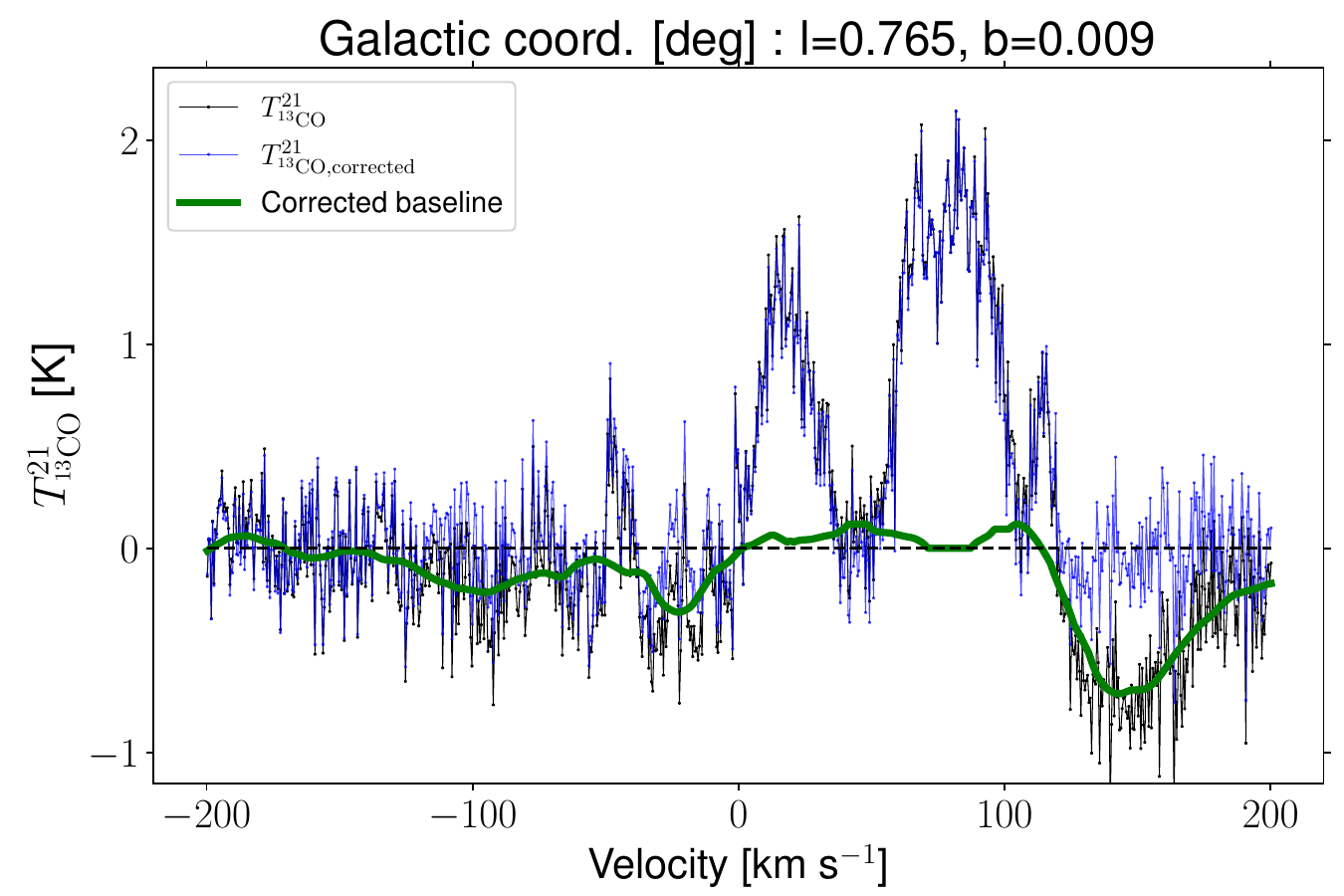}
    \includegraphics[width=0.8\columnwidth]{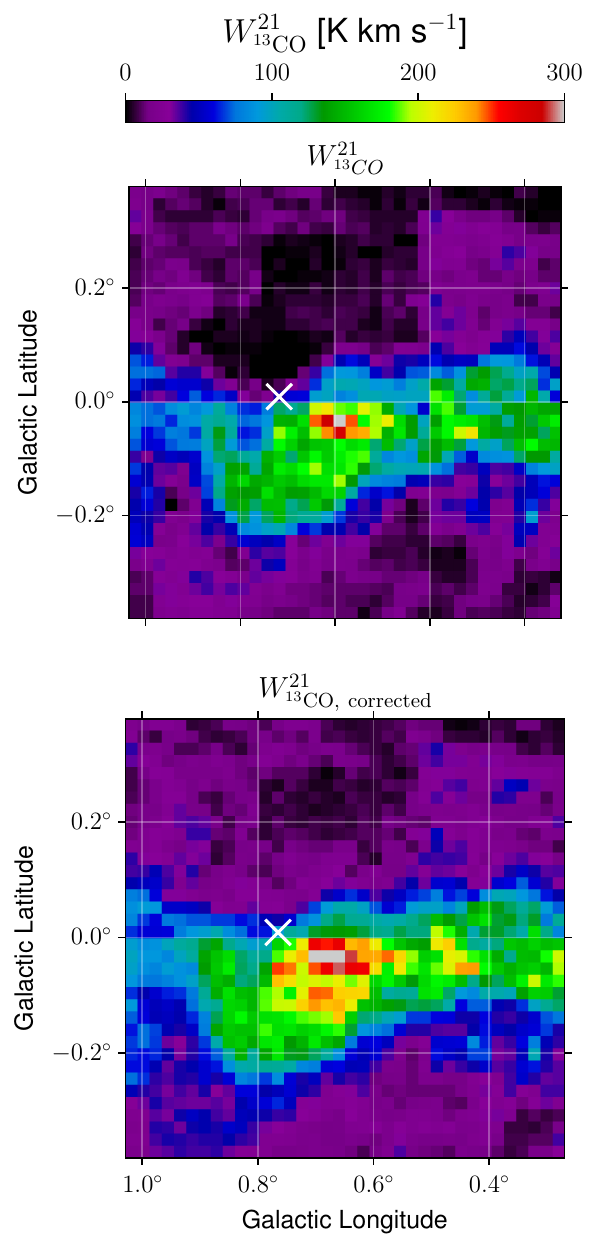}
    \caption{Top: Brightness temperature profile (black) for the line of sight indicated by the white cross in the middle and bottom panels, the running mean (green) and the corrected brightness temperature (blue) computed as the difference between the original profile and the running mean. Middle: Original integrated brightness temperature of \thcoii from APEX, denoted as \wthcoii. Bottom: Integrated brightness temperature of \thcoii after the correction for the baseline, denoted as $W_{^{13} \rm CO,\ corrected}^{21}$.  }
    \label{fig:APEX}
\end{figure}

\subsubsection{\hi absorption correction}\label{sec:hi_abs}

Several strong continuum sources in the GC region, when viewed in absorption, show up as negative sources in the continuum-subtracted \hi line
cubes. We choose to discard the data from the pixels affected by absorption, and try instead to reconstruct their potential signal using the information from the nearby pixels.

First, we identify the regions with a sharp decrease in brightness temperature, $T_{\rm B}$.
For each velocity channel, we compute the $T_{\rm B}(l,b)$ image gradient using the Sobel operator and we apply a hysteresis filtering\footnote{\href{https://scikit-image.org/docs/stable/api/skimage.filters.html\#skimage.filters.sobel}{Sobel filter} and \href{https://scikit-image.org/docs/stable/auto_examples/filters/plot_hysteresis.html}{hysteresis thresholding} in scikit-image documentation \citep{scikit-image}.} with low and high thresholds at 95 and 99.9 percentiles in image gradient, respectively. This filter masks the pixels above the high threshold and the pixels between the two thresholds that can be continuously connected to a pixel above the high threshold. We also impose values in brightness temperature to be lower than its 90 percentile, or lower than -5 K, in order to avoid selecting sharp positive variations in the signal.
The masked pixels are considered to be affected by absorption. 

For large optical depth, the brightness temperature of the \hi line tends to saturate. As this is the case in most directions toward the GC \citep{2017MNRAS.468.4030S}, we can expect the masked pixels to have a similar value than the maximum one seen in the nearby pixels. 
Then, for each latitude, we get the maximum value in brightness temperature within 10 pixels ($~0.2\degree$) in longitude around the masked pixels, and we fill them with this value. The procedure is illustrated in Fig.~\ref{fig:NHImethod} in the appendix. 
We also considered using the median value instead of the maximum, but we adopted the maximal correction because the map with the median correction still contains absorption artefact (as holes and stripes) toward the GC as shown in Fig.~\ref{fig:NHImaps} in the appendix. 

After this correction, the \hi data cubes are downsampled to match the common spatial resolution used for the rest of the analysis. The relation used to compute the hydrogen column density is given in section \ref{sec:nhi}.

\subsubsection{CO baseline correction}\label{sec:baseline}

The baseline of the \co line emission profiles in the public datasets provided by the 45m NRO and the APEX observatories are found to be shifted from zero toward negative values over large ranges of velocities in several lines of sight.
To correct this deviation, a new baseline is computed for each velocity channel as a running mean in a $60 \ \rm{km \ s^{-1}}$ window.
While computing the running mean, region with significant signal (larger than standard deviation) are clipped to zero.
An example profile can be found in the top panel of Fig.~\ref{fig:APEX} with the running mean shown as the green line. The example profile correspond to the line of sight indicated by the white cross in the middle and bottom panels. 

The middle and bottom panels of Fig.~\ref{fig:APEX} illustrate the integrated \co line emission before and after the baseline correction, respectively. After the baseline correction, one can notice that the integrated intensity over velocity is, on average, larger and that many more gas structures can be appreciated while the zero-filled rectangular patches vanish. This is especially noticeable in the region near the Sgr~B2 cloud (around $l = 0.7\degree, b=0\degree$) and in the faint regions north and south of Sgr~B2.

For both surveys, this correction strongly attenuates the disparities in sensitivity across different patches (likely associated with different observation phases). In the end, this effectively improves their average sensitivity. 

\subsection{Dust emission}

The spectral energy distribution of the emission from dust grains, in the optically thin limit, can be described as:
\begin{equation}
    I_\nu = \tau_\nu B_\nu(T),
\end{equation} where $I_\nu$ is the dust specific intensity, $\tau_\nu$ is the optical depth, and $B_\nu(T)$ is the Planck function for dust at temperature $T$.

In \citet{2014A&A...571A..11P}, due to the limited number of photometric bands available (\textit{Planck} $353$, $545$, and $857$ GHz and IRAS $100$ $\upmu$m bands) a single modified blackbody (MBB) model was adopted for each sight line, thus assuming that the integral emission of dust along the line can be fitted as a single-component dust mixture with an average temperature. The optical depth $\tau_{\nu}$ can be expressed as a power-law $\tau_{\nu} = \tau_{\nu_0} (\nu / \nu_0)^{\beta}$, where $\nu_0$ is the reference frequency at which $\tau_{\nu_0}$ is estimated. Then, the single MBB can be parametrized as:
\begin{equation}
    I_\nu = \tau_{\nu_0} \bigg( \frac{\nu}{\nu_0} \bigg)^{\beta} B_\nu(T).
    \label{eq:tau_1comp}
\end{equation}

\begin{figure}
    \centering
    \includegraphics[width=\columnwidth, trim={0.65cm 0.7cm 1.2cm 0.65cm},clip]{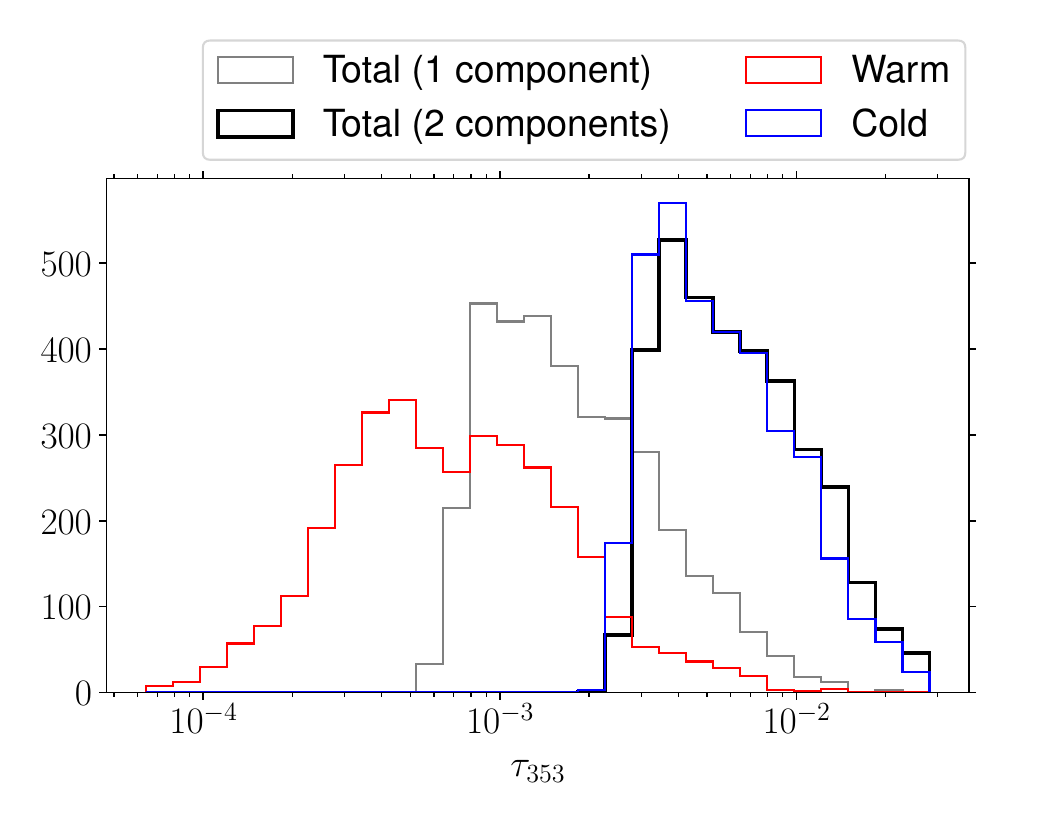}
    \caption{Dust optical depth distributions. The grey line corresponds to the result obtained by fitting a single dust component model and the others to the fit result for the two dust components model.
    }
    \label{fig:tau_hist}
\end{figure}

Several authors suggested more sophisticated methods using two dust components. \citet{1995ApJ...451..188R} fitted the COBE/FIRAS observations using single and two-component models and showed that two-component models were significantly better than single-component models. \citet{1999ApJ...524..867F} tried to extrapolate the dust emission at 100 $\upmu$m and Cosmic Microwave Background radiation from FIRAS observations and showed that single-component models were unable to accurately match the FIRAS and DIRBE spectrum at both the Wien and Rayleigh-Jeans regimes. Thus, they adopted an emission model comprised of two MBBs, each at a different dust temperature and emissivity power-law index, representing two distinct dust grain species. Later on, \citet{2015ApJ...798...88M} applied \citet{1999ApJ...524..867F}'s model to the \textit{Planck}-HFI and the joint \textit{Planck}/DIRBE maps fitting the $100$-$3000$ GHz emission bands, and gave accurate predictions in the $100$-$217$ GHz range. 

The two-component MBB model can be expressed as:
\begin{equation}
    I_\nu = \tau_{\nu_0}^{\rm warm} \bigg( \frac{\nu}{\nu_0} \bigg)^{\beta_{\rm warm}} B_\nu(T_{\rm warm}) + \tau_{\nu_0}^{\rm cold} \bigg( \frac{\nu}{\nu_0} \bigg)^{\beta_{\rm cold}} B_\nu(T_{\rm cold}).
    \label{eq:tau_2comp}
\end{equation} The reference frequency is taken as $\nu_0= 353$ GHz. We adopt a spatially fixed $\beta_{\rm warm}$, as suggested in \citet{2015ApJ...798...88M} and with the best-fit value of 2.7 as found in \citet{1999ApJ...524..867F}. The free parameters $T_{\rm warm}$, $ \tau_{\nu_0}^{\rm warm}$, $\beta_{\rm cold}$, $T_{\rm cold}$, $ \tau_{\nu_0}^{\rm cold}$, are fitted to the data by a $\chi^2$ minimization method.
The two-component MBB fit is performed using the observations from the \textit{Herschel} PACS and SPIRE instruments, the APEX/LABOCA Telescope, and the \textit{Planck}-HFI instrument\footnote{Three of the emission bands are contaminated by \co molecular transition line emission: 100, 217 and 353 GHz, and have been corrected using the same method as \citet{2015ApJ...798...88M}.} (frequencies and references are given in Table \ref{tab:tracers}). This gives us a total of 12 data points per line of sight. However, due to the coarser resolution of \textit{Planck}-HFI, worse than our target resolution of $\sim0.02 \degree$, we divide the fitting process into two steps. First, we fit all free parameters, using the data from all four instruments, at the resolution of \textit{Planck}-HFI maps. Then, we refit the optical depths $\tau_{\nu_0, 1}$ and $\tau_{\nu_0, 2}$ for both warm and cold dust components with only the PACS, SPIRE and APEX/LABOCA data (6 data points per sight line) at $\sim0.02 \degree$ resolution, fixing the parameters $\beta_{obs,2}$, $T_{obs, 1}$ and $T_{obs, 2}$ to the values obtained at lower resolution in the first step. Results are shown in Fig.~\ref{fig:dust_2comp_results} in appendix. 

For comparison, we also derive maps of $\tau$, $\beta$, and $T$ for the single-component MMB model described by Eq. \ref{eq:tau_1comp} fitting only to the \textit{Herschel} and APEX/LABOCA data. Results are shown in Fig.~\ref{fig:dust_1comp_results} in appendix. The distribution of the average optical depths obtained with the two-component and the single-component models are shown in Fig.~\ref{fig:tau_hist}.

\begin{figure*}
    \centering
    \includegraphics[width=\textwidth, trim={0cm 8cm 0cm 0cm},clip]{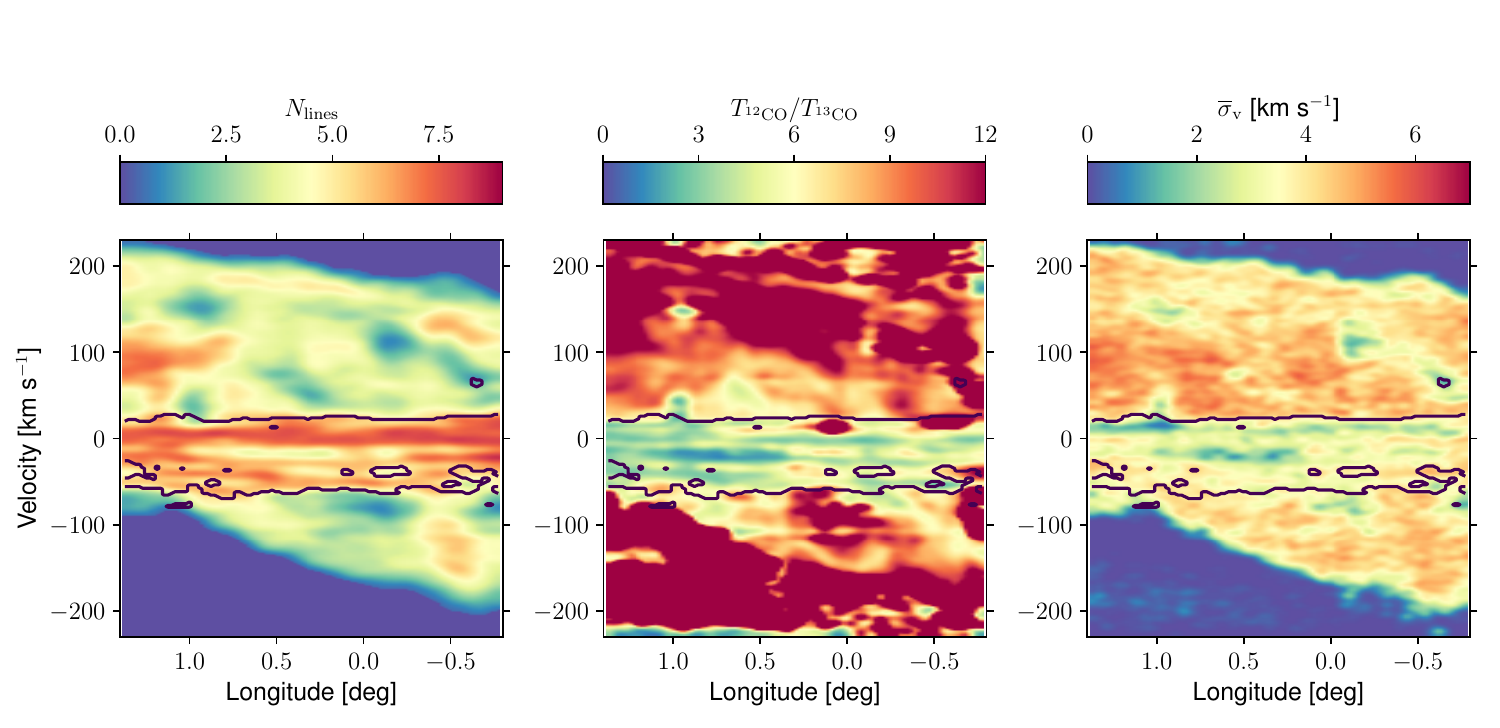}
    \caption{From left to right: longitude-velocity diagrams of the number of \twco lines, the brightness temperature ratio of \twco to \thco, and the \twco line width.  For each (l,v) pixel, the number of lines is summed over latitudes while the two other quantities are averaged. The maps have been smoothed with a Gaussian kernel of 2 pixels in standard deviation for display. The contours give the outline of the two components separated by hierarchical clustering using these three parameters maps as input. Pixels outside of the contour are associated with the CMZ, and those inside are associated with the disk, if they also verify the other conditions described in the text (see Sect. \ref{sec:disk_sep}).
    }
    \label{fig:lv_CMZextraction}
\end{figure*}

\section{Methods} \label{sec:method}
 
Our goal is to create a map of the total hydrogen column density in the CMZ by combining \hi data and \co isotopologue observations. In this section, we explain the process of decomposing the velocity profiles of \hi and \co data into individual emission lines, which allows us to distinguish the CMZ from the gas in the disk. By using simulation predictions and empirical corrections, \xco factors are derived for each emission line.

 \subsection{Line profile decomposition}\label{sec:decomposition} 
 
For each longitude and latitude, the brightness temperature profile as a function of velocity is decomposed in multiple line contributions following the method of \citet{2015A&A...582A..31P, 2017A&A...601A..78R}. Detecting lines consists of finding the velocities at which there are local minima in the second derivative of the brightness temperature profile. At these velocities, the brightness temperature presents a local maximum or a change of slope, indicating the presence of a line. 

A Gaussian smoothing is applied to the profile before line detection to reduce fluctuations due to noise. Then, the second derivative of the smoothed profile is computed. Only lines found as a local minima of the second derivative within a range $\Delta v$ (minimal line separation) and above $\sigma_{Thr}$ times the threshold $T_{\rm Thr}$ are considered. $T_{\rm Thr}$ is defined as the standard deviation of the profile computed after clipping $T_{\rm B}$ values below the median of the full profile plus the median absolute deviation.  
The parameters used for the line detection in the emission profiles are listed in Table \ref{tab:line_parameters} in the Appendix. 

Each detected line is then modelled as a pseudo-Voigt function:
$pV = \eta L + (1-\eta) G$ where $0 < \eta < 1$ is the form factor, $G$ a Gaussian and $L$ a Lorentzian distribution. These two functions are expressed as $ G = h\exp{\left(\frac{v-v_0}{\sigma}\right)^2}$
    and $ L = h / \left(1 + \left(\frac{v-v_0}{\sigma}\right)^2\right)$
where $h$ is the height of the line, $\sigma$ its width and $v_0$ its central velocity. $v_0$ is constrained into the range $[v_* - dv; v_* + dv]$ with $dv$ being the velocity resolution of the survey, and $v_*$ the velocity of the line obtained previously in the detection step.

The sum of their contributions is fitted to match the observed brightness temperature profile. So for each detected line, the best-fit parameters for $\eta$, $h$, $\sigma$, and $v_0$ are found through a $\chi^2$ minimization method.

In order to preserve the observed photometry, we calculate the residuals between the observed and fitted spectra in each velocity channel, and we redistribute them among the fitted lines proportionally to their intensity in that channel. Finally, we compute the integrated intensity for each line by integrating the pseudo-Voigt function along the entire velocity range.

\subsection{Disk component separation}\label{sec:disk_sep} 

The contamination from the Galactic disk to the CMZ CO emission is not negligible. We rely on the differences in gas properties between the CMZ and the rest of the Galaxy to separate these two components.

Higher turbulence levels and shear in the CMZ yield larger velocity dispersion within a cloud and from cloud-to-cloud compared to the disk. Hence, emission lines in the CMZ are more spread apart in velocity. For any given line of sight, the number of emission lines, $N_{\rm lin}$, is higher in the Galactic disk than in the CMZ, while the average line width $\bar{\sigma}_v$ is higher in the CMZ.
Moreover, a higher turbulence tends to decrease the optical thickness of the \twcoi emission lines, increasing their maximal brightness temperatures $T_{^{12} \rm{CO}}$. As the \thco emission lines are less optically thick, they are less affected. Therefore, the ratio of the brightness temperatures $T_{^{12} \rm{CO}}/T_{^{13} \rm{CO}}$ is higher in the CMZ compared to the Galactic disk.

We first separate the disk gas from the CMZ in the \thcoi brightness map by considering, $N_{\rm lin}$, $\bar{\sigma}_v$ and $T_{^{12} \rm{CO}}/T_{^{13} \rm{CO}}$. For each longitude and velocity value we compute the sum of $N_{\rm lin, ^{12} \rm{CO}}$, and the average of $T_{^{12} \rm{CO}}/T_{^{13} \rm{CO}}$ and $\bar{\sigma}_{v, ^{12} \rm{CO}}$ over the small latitude range of our map. Theses parameters maps as a function of $(l,v)$ are shown in Fig.~\ref{fig:lv_CMZextraction}.
We then use an inductive clustering approach to isolate two clusters in this parameter space.
First, we perform an agglomerative clustering\footnote{\href{https://scikit-learn.org/stable/auto_examples/cluster/plot_inductive_clustering.html}{InductiveClustering} and \href{https://scikit-learn.org/stable/modules/generated/sklearn.cluster.AgglomerativeClustering.html}{AgglomerativeClustering} documentation from scikit-learn \citep{scikit-learn}.} using only 20\% of the $(l,v)$ pixels, randomly chosen. Then, the obtained clusters are used to train a support vector classifier (SVC)\footnote{used with \texttt{probability = True, gamma = 1, C = 100}, see \href{https://scikit-learn.org/stable/modules/generated/sklearn.svm.SVC.html}{SVC} documentation.}. This classifier is applied to the full data, giving us a boolean mask in longitude and velocity.

Finally, we add two conditions on individual line parameters for the disk component: (i) upper limits of 100 and 5 $\mathrm{K}\,\mathrm{km}\,\mathrm{s}^{-1}$ on the integrated intensity of \twcoi and \thcoii, respectively. (ii) upper limit of $\pm80~\mathrm{km}\,\mathrm{s}^{-1}$ for the line velocity. These values are chosen because higher values will be reached mostly toward CMZ clouds. The same mask is used for \twcoi, \thcoii and \hi data.
The result of the component separation for the final column density maps is presented in Sect. \ref{sec:res_maps}.

\subsection{Hydrogen column density estimates}\label{sec:nh_estimate} 
 \subsubsection{Atomic phase: \nhi}\label{sec:nhi}

The column density of atomic hydrogen, \nhi, is given by:
\begin{equation}
    N_{\rm HI}(l,b) = -X_{\rm HI} T_{\rm S} \int \mathrm{ln}\left(1 - \frac{T_{\rm B}(l,b,v)}{T_{\rm S}}\right)\,dv
\label{eq:nhi}
\end{equation} with $X_{\rm HI}=1.82\times10^{18}~\mathrm{cm^{-2}} (\mathrm{K}\,\mathrm{km}\,\mathrm{s}^{-1})^{-1}$ the conversion factor. $T_{\rm B}$ is the  brightness temperature obtained after continuum subtraction \citep{2012ApJS..199...12M} and after applying the \hi absorption correction described in section \ref{sec:hi_abs}. Thanks to the components separation (section \ref{sec:disk_sep}) we can split the lines associated to the disk and the CMZ so $T_{\rm B}=T_{\rm B}^{\rm disk}+T_{\rm B}^{\rm CMZ}$. Then \nhi is computed independently for each component. $T_{\rm S}=146.2\pm16.1~\mathrm{K}$ is the spin temperature derived by measuring saturated brightness in the radial-velocity degenerate regions toward the GC given by \citet{2017MNRAS.468.4030S}. This value is valid for the gas in the disk but not necessarily for the gas in CMZ since it is not well-resolved in \hi data. We assume the same $T_{\rm S}$ value for the disk and CMZ components. The effect of the spin temperature choice is further discussed in Sect. \ref{sec:mass}.

 \subsubsection{Molecular phase: \nhd}\label{sec:nhd}

 The \co line emission is usually scaled into molecular hydrogen column density as: 
\begin{equation}
    N_{\rm H_2} = X_{\rm CO} W_{\rm CO}.
\label{eq:nh2}
\end{equation} 
The \xco factor can vary within a cloud and across the Galaxy depending on several environmental parameters. Its variation due to the surface density, the metallicity, the far UV radiation field strength and the CR ionization rate has been studied by \citet{2020ApJ...903..142G} using 3D magnetohydrodynamic simulations, for a range of observational beam sizes. We use their results to describe the evolution of \xco with metallicity, $Z$ (in units of $Z_\odot$), resolved scale, $r$, and integrated intensity of each line decomposed in velocity, $W$: 
\begin{align}
X_{^{12}\rm CO}^{10} &= f(W_{^{12}\rm CO}^{10}) {W_{^{12}\rm CO}^{10}}^{0.19\log(r)} Z^{-0.8} r^{-0.25} \label{eq:xco12} \\
X_{^{12}\rm CO}^{21} &= g(W_{^{12}\rm CO}^{21})  {W_{^{12}\rm CO}^{21}}^{0.34\log(r)} Z^{-0.5} r^{-0.41}
\end{align}
where $f$ and $g$ are derived from the mean values given by the simulation (see yellow points in Figure 10 and 15 of \citet{2020ApJ...903..142G} for \twcoi and \twcoii, respectively). As these reference values were computed for  \mbox{$Z=1$} and  \mbox{$r=2$} pc we correct them for the resolved scale, so \mbox{$f = X_{^{12}\rm CO, figure}^{10}(W_{^{12}\rm CO}^{10}, Z=1, r=2 \, \rm pc) / ({W_{^{12}\rm CO}^{10}}^{0.19\log(2)}2^{-0.25})$}, and a similar correction applies to $g$.

\begin{figure}
    \centering
    \includegraphics[width=\columnwidth, trim={0.7cm 0.7cm 0.5cm 7.5cm},clip]{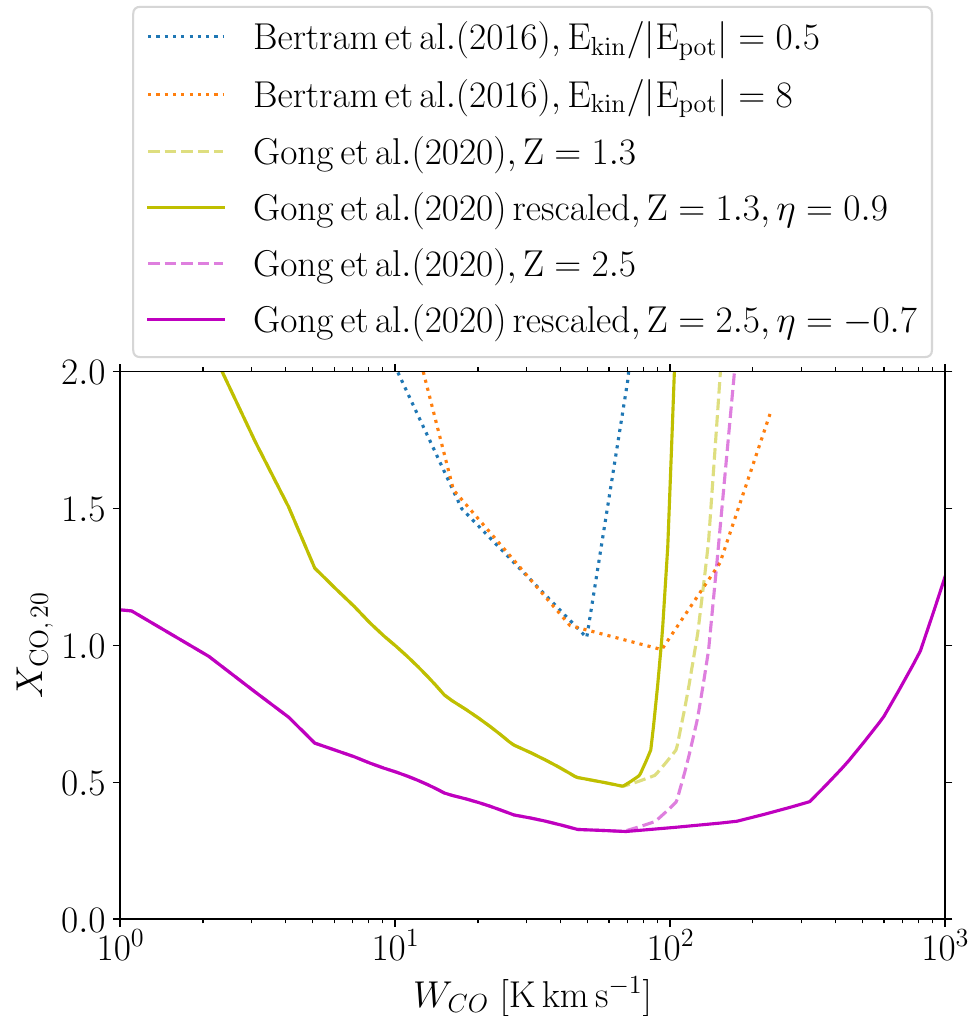}
    \caption{Mean trend for the \xco factor evolution as a function of \wco as predicted by simulations, adapted from Figure 6 of \citet{2016MNRAS.455.3763B} and Figure 10 of \citep{2020ApJ...903..142G}. The effect of the turbulence is taken into account in \citet{2016MNRAS.455.3763B} by varying the ratio of kinetic and potential energies (dotted curves). The dashed curves correspond to the reference functions from \citet{2020ApJ...903..142G}, noted $f($\wtwcoi$)$ in Eq. \ref{eq:xco12}. A power-law correction of index $\eta$ can be applied to this curve in order to mimic the effect of the turbulence, as proposed in Eq. \ref{eq:xco12_eta}.}
    \label{fig:xco_vs_wco_models}
\end{figure}

Another key parameter at play in the CMZ is the increased level of turbulence, which yields a larger velocity dispersion, effectively reducing the optical thickness of the \co lines compared to the gas in the disk. The simulation of \citet{2016MNRAS.455.3763B} considers different turbulence levels by varying the virial parameter, defined as the ratio of kinetic and potential energies. We notice that variations of this parameter mostly modify the shape of the \xco as a function of \wco at high \wco values (after the minimum in \xco), while at low \wco the average trend is unchanged \citep[and very similar to the one predicted by][]{2020ApJ...903..142G}.
So we propose that the effect of the turbulence can be taken into account empirically by adding a power-law correction to Eq. \ref{eq:xco12} such as :
\begin{align}
X_{^{12}\rm CO}^{10} &= f(W_{^{12}\rm CO}^{10}) \left(\frac{W_{^{12}\rm CO}^{10}}{W_{^{12}\rm CO, \rm Xmin}^{10}}\right)^\eta {W_{^{12}\rm CO}^{10}}^{0.19\log(r)} Z^{-0.8} r^{-0.25} \label{eq:xco12_eta}
\end{align}
The comparison of the \xco predictions from these two simulations is shown in Fig.~\ref{fig:xco_vs_wco_models} and the result of this correction on the \xco evolution with \wco is further discussed in Sect. \ref{sec:res_fit}.

\begin{figure}
    \centering
    \includegraphics[width=\columnwidth, trim={0.5cm 0.0cm 0.5cm 0.5cm}, clip]{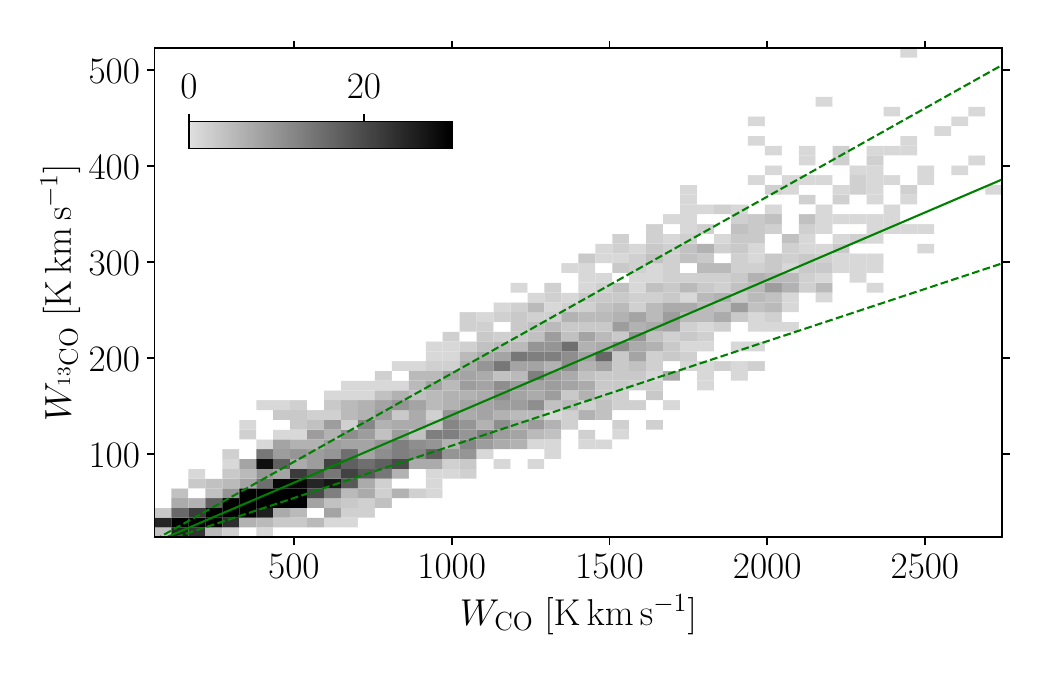}
    \includegraphics[width=\columnwidth, trim={0.5cm 0.5cm 0.5cm 0.2cm}, clip]{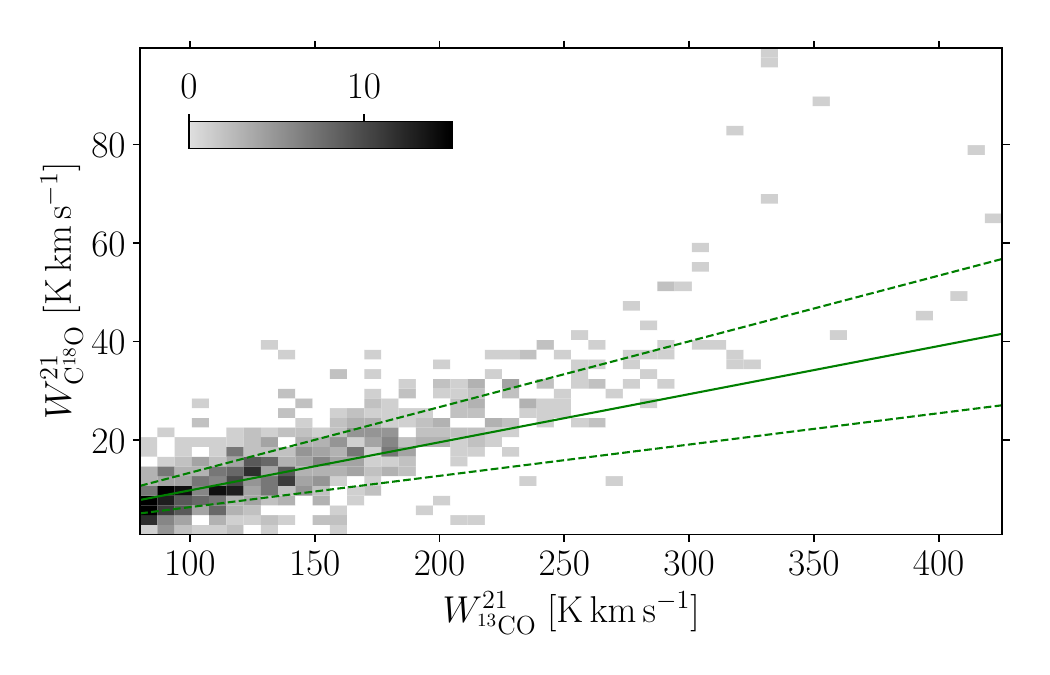}

    \caption{Top: Integrated line intensities of \thco versus \twco isotopes for the $(1-0)$ transition.
Bottom: Integrated \co line intensities of the \eico versus \thco isotopes for the $(2-1)$ transition.
Green full lines correspond to linear relation for which the slope is the median \wco ratio; dashed lines give $1\sigma$ errors from the 16 and 84 percentiles.
}
    \label{fig:wcos}
\end{figure}

Note that these simulations provide only the trend for \wtwcoii while we have \wtwcoi observations. So we derive a median conversion factor $ \overline{R}_{1312}$ between \twcoi and \thcoi isotopes, for which we have both observations available (and they correlate well enough to use only the median as scaling factor, see top panel of Fig.~\ref{fig:wcos}). This assumes that the $\overline{R}_{1312}$ value is similar for the two transitions. 
The observations by \citet{Nishimura_2015} in Orion clouds support this assumption, finding moreover similar value of $ \overline{R}_{1312}$ as the one we calculate for the GC. This translates to:
\begin{align}
    \overline{R}_{1312} &= \rm{median}(W_{^{13}\rm CO}^{10} / W_{^{12}\rm CO}^{10}),  \\
    W_{^{12}\rm CO}^{21} &= W_{^{13}\rm CO}^{21} / \overline{R}_{1312}.
\end{align}

In the simulations of \citet{2016MNRAS.455.3763B}, there is no evidence of saturation in the \thcoi transition, so the \xco is nearly constant at large \wco, and the turbulence parameter has little impact. So, for rarer isotopes, we don't introduce the $\eta$ parameter. We check the correlation of \thcoii with \eicoii (see bottom panel of Fig.~\ref{fig:wcos}) and find no evidence of saturation except toward the core of Sgr~B2 and the GC. In these regions, we use a similar approach to \citet{2017MNRAS.471.2523Y} (see the text describing Figure 3 in their paper) and derive a saturation-corrected \wthcoii using as substitute \weicoii scaled by the \wco ratio found in the regime where a linear relation exists between the two isotopes.
So for $80<$\wthcoii$<200$ \wcounit :
\begin{align}
    \overline{R}_{1813} &= \rm{median}(W_{\rm C^{18}O}^{21} / W_{^{13}\rm CO}^{21}),
\end{align}
and if \wthcoii $> W_{^{13}\rm CO, \, observed}^{21}> 200$ \wcounit :
\begin{align}
    W_{^{13}\rm CO}^{21} &= W_{\rm C^{18}O}^{21} / \overline{R}_{1813}.
\end{align} 

Finally, we re-scale the trend predicted by the simulations by fitting the Z and $\eta$ parameters such that the $\rm H_2$ column density maps derived from \twcoi lines matches the ones obtained from \thcoii lines :
\begin{align}
    N_{\rm H_2}(l,b) &= \sum_i^{\rm{n_{lines}}} X_{^{12}\rm CO, i}^{10}(W_{^{12}\rm CO, i}^{10}, Z,r,\eta) \, W_{^{12}\rm CO, i}^{10}(l,b) \label{eq:nh2_12co} \\
    N_{\rm H_2}(l,b) &=  \sum_i^{\rm{n_{lines}}} X_{^{12}\rm CO, i}^{21}(W_{^{12}\rm CO, i}^{21}, Z,r) \, W_{^{12}\rm CO, i}^{21}(l,b) \label{eq:nh2_12co21} 
\end{align}
where $X_{\rm CO, i}$ and $W_{\rm CO, i}$ are the quantities derived for each line decomposed in velocity.
The difference between the two \nhd estimates is minimized using a least squared method. The CMZ and disk parameters can be fitted independently thanks to the component separation. The best-fit parameters are given in Sect.~\ref{sec:res_fit}.

\begin{figure*}
    \centering
    \includegraphics[width=.9\textwidth, trim={0.95cm 0.5cm 0.5cm 0.3cm},clip]{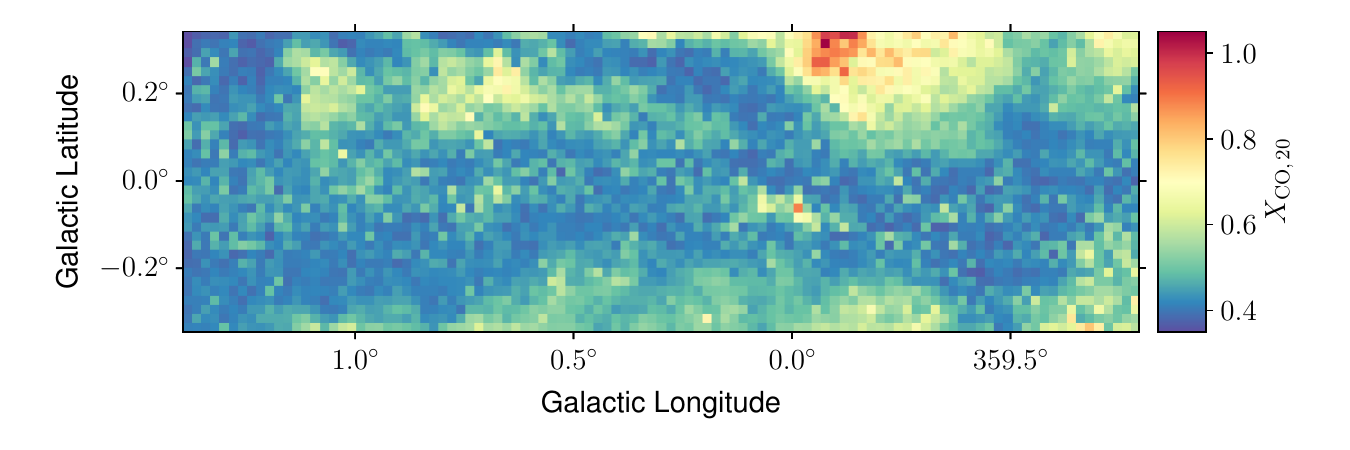}
    \caption{Map of \xcotwenty, the CO-to-\hd conversion factor in units of  10$^{20}$ cm$^{-2}$K$^{-1}$km$^{-1}$s, averaged for each pixels. }
    \label{fig:xco_map}
\end{figure*}

\section{Results and discussions on \co lines properties}\label{sec:res_co}

As detailed in the previous section, our \nhd estimates require two \co isotope ratios and several environmental parameters in order to derive \xco. These quantities and associated errors are given in the following.

 \subsection{Isotope integrated intensity ratios}\label{sec:res_iso}

The integrated intensities of the $(1-0)$ transition for the \twco and \thco isotopes are well-correlated, with a Pearson correlation coefficient of 94\%. This can be seen in the 2D histograms in the top panel of Fig.~\ref{fig:wcos}. Over the whole range of \wco, we find a median ratio $ \overline{R}_{1312} = 0.14 ^{+0.04} _{-0.03}$ (the positive, and negative errors are derived from the 16, and 84 percentiles, respectively). This value is slightly larger but still compatible with the mean value $0.12 \pm 0.01$ reported in an earlier study using the same data \citep{2019PASJ...71S..19T}, even if we do not mask out the disk component, and despite the additional baseline correction we applied (see Sect. \ref{sec:baseline}). 

The integrated intensities of the $(2-1)$ transition for the \eico and \thco isotopes are shown in the bottom panel of Fig.~\ref{fig:wcos}. Over the whole range of \wco, the correlation is only of 77\%. Indeed we note departures from linearity: (i) at large \wco, in correspondence with the core of Sgr~B2 and Sgr~A, where the \thco intensity saturates; (ii) at low \wco where there is not enough sensitivity, especially to detect the \eico emission, and lower signal-to-noise ratio. In the intermediate range $80<$\wthcoii$<200$ \wcounit, where we have a more linear relation, we measure a median ratio $ \overline{R}_{1813} = 0.10 ^{+0.04} _{-0.03}$. 

The good correlation between \twcoi and \thcoi over a large range of \wco values and the absence of saturation, except toward the core of Sgr~B2 and Sgr~A, hints that the lines' optical thickness does not rise steeply and remains rather low in most directions toward the CMZ. By adopting the local thermal equilibrium (LTE) approximation, \citet{2019PASJ...71S..19T} estimated that the \twcoi line optical depths
distribution toward the CMZ peaks at 3, and values larger than 10 originate predominantly from regions with large contamination from the disk.
Earlier studies as \citep{1998A&A...331..959D} have already shown that the optical thickness is much lower in the CMZ ($\tau \lesssim 2$ for \wtwco / \weico $> 60$) than in the disk ($\tau \geq 10$), except toward few dense cores such as Sgr~B2.
\citet{SAWADA1999985} also reported lower \twcoi optical depths in the GC compared to the disk, which they explained as follows: (i) even for similar densities, higher temperatures excite many molecules to high-J states, reducing optical depths for lower-J transitions. (ii) larger velocity dispersions, caused by strong external pressure and tidal shear in steep gravitational potential, further reduce optical depths.

\subsection{\xco fit parameters}\label{sec:res_fit}

The parameters controlling the evolution of \xco that cannot be taken from the literature are estimated from the fit, minimizing the difference of the \nhd estimates from the two isotopes using a least-squared method. The systematic uncertainties on the model are estimated as a fractional value of predicted \nhd, such that the reduced \chisq reaches unity. This condition is verified considering an error of 10\% for the estimate derived from \thcoii, and 33\% for the one derived from \twcoi, the latter is larger due to the line saturation. 

Thanks to the component separation, parameters for both the CMZ and the disk can be fitted.
The resolved scale for the CMZ component is $r_{\rm CMZ}=2.8$ pc, given the angular resolution of 0.02$\degree$ and a distance to GC of 8.18~kpc \citep{2019A&A...625L..10G}. 
The metallicity in the central parsec of the Milky Way was measured by \citet{2015ApJ...809..143D} from K-band spectroscopy of 83 late-type giants stars, and they found a mean value of $\rm{[M/H]}= +0.4\pm0.4$~dex. Other studies support that the GC contains mainly a metal-rich population of stars with a mean metallicity of +0.2~dex \citet{2019A&A...627A.152S} or +0.3~dex \citet{2018MNRAS.478.4374N}. 
It should be noted that the stellar metallicity does not always match the gas-phase metallicity, at the centres of massive Galaxies ($> 10^{10} \; \rm{M_{\odot}}$) the stellar metallicity tends to be slightly higher in average \citep{2022MNRAS.510..320F, 2024A&A...690A..13Z}, but the differences remain lower than the uncertainties for the centre of our Galaxy. If the metallicity is fixed to $Z_{\rm CMZ}=2.5$ according to the higher mean value reported in the literature, then the index obtained from the fit is $\eta_{\rm CMZ}=-0.71\pm0.01$. This matches the expectations for the CMZ as $\eta$ values lower than zero correspond to an increased level of turbulence compared to the reference value ($\eta=0$).
If the metallicity is left free, we find $Z_{\rm CMZ} = 1.92 \pm 0.02$ ($\sim +0.3$~dex) and $\eta_{\rm CMZ}=-1\pm0.1$, which is still compatible with the measured metallicities and the $\eta<0$ expectation for the CMZ. However, in that case, there is some degeneracy between the two parameters, and the quality of the fit is worse at large \nhd, so in the following, we consider the results from the fit at fixed metallicity shown in Fig.~\ref{fig:xco_vs_wco_models}.

\begin{figure}
    \centering
    \includegraphics[width=\columnwidth, trim={0.5cm 0.5cm 0.5cm 0.5cm}, clip]{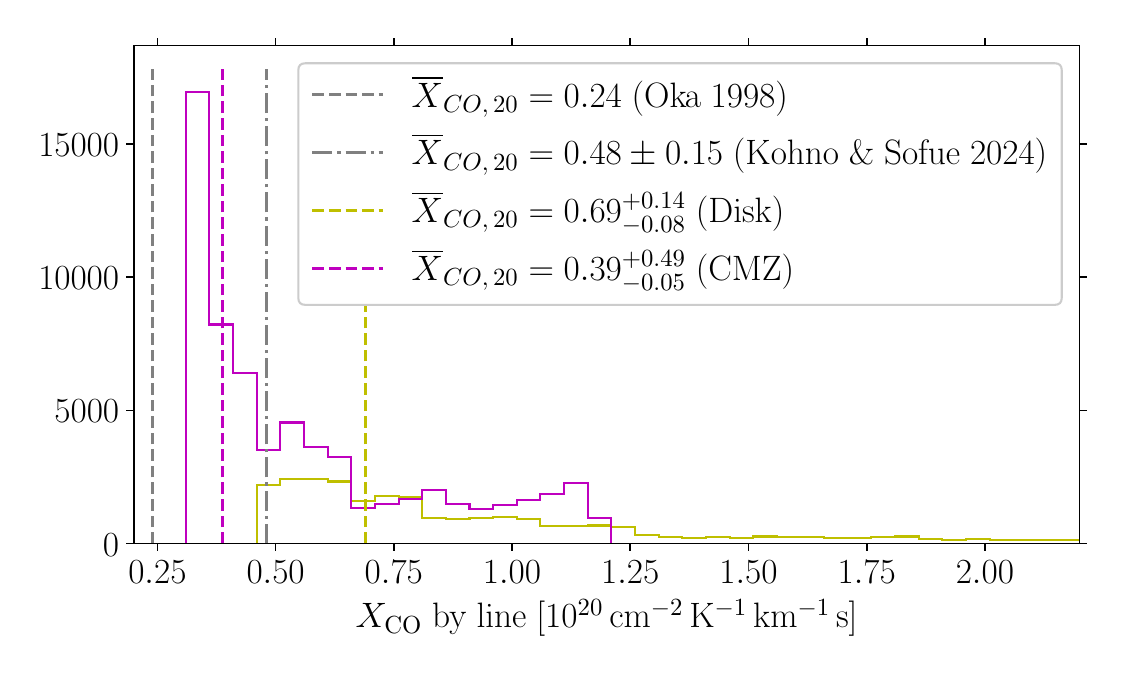}
    \caption{Histogram of the \xco factor derived for each fitted line. The grey dashed, and dashed-dotted lines correspond to the average measurements of \citet{1998ApJ...493..730O}, and \citet{2024PASJ...76..579K}, respectively. The magenta, and yellow lines corresponds to the median of \xco maps for the CMZ, and disk components. Negative and positive errors are derived from the 16 and 84 percentiles of the map.}
    \label{fig:xco_hist}
\end{figure}

For the disk component, we only considered mean parameters for simplicity, but this approximation still offers a first-order correction compared to studies that do not consider its different properties. The impact of this approximation is limited because most of the disk gas toward the CMZ belongs to the inner regions that have much larger densities than the outer regions, so the average values are driven by the gas in a limited range of distances with rather similar properties. Moreover, given the moderate contribution of the molecular gas in the disk toward the CMZ, this approximation won't introduce a severe bias on the total column density derived. Nevertheless the fitted parameters for the disk component should be considered as nuisance parameters, they are given only for completeness, and not for physical interpretations. The parameters fitted are a mean index $\overline{\eta}_{\rm disk}=0.9\pm0.1$, and a mean distance, $\overline{d}_{\rm disk} = 2300\pm400$~pc. This distance yields a mean resolved scale of $r_{\rm disk}=0.8\pm0.1$ pc. A mean metallicity $\overline{Z}_{\rm disk}=1.3\pm0.1$ is derived from the fitted distance considering a gradient of $-0.047$~dex/kpc as measured by \citet{2018A&A...618A.160L} using Gaia DR2 data. Even if we don't want to interpret these values, we can still note that the fitted trend goes in the right direction with lower metallicity and larger $\eta$ (so lower turbulence) in the disk component compared to the CMZ.

The statistical uncertainties on the parameters controlling \xco, derived for the CMZ regions as a whole, are only a few percent. By contrast, the systematic uncertainty on the \nhd estimates is much higher with values of 10-30\%. This could be explained by variations of the isotopes ratio, as indeed their dispersion is of similar order (simulations show large uncertainties as well). Moreover, variations of the metallicity across the CMZ could contribute to the variations of the \xco factor. Furthermore, it is unclear how the turbulence parameter would vary depending on the position of the cloud and its dynamic with respect to the GC.

\subsection{Xco variations}\label{sec:xco}

\xco is derived for each detected line considering  Eq. \ref{eq:xco12_eta} with the fitted parameters given in Sect. \ref{sec:res_fit}.
The map in Fig.~\ref{fig:xco_map} gives the average \xco in each direction. Each pixel value in the \xco map is the \wco-weighted average of the \xco over all the detected lines belonging to a given component in that direction. The overall distributions of \xco per detected line for each component are shown in Fig.~\ref{fig:xco_hist}. 

For the CMZ component, the \xco values are in the range \mbox{$(0.32$ -- $1.37) \, \times$ \xcounit} for each line, but the distribution is very asymmetric and very skewed toward the lower bound. This directly comes from the fact that most detected lines have \wco between few tens and few hundreds of \wcounit, and the $\rm{X_{CO}}(W_{\rm CO})$ profile is nearly flat close to its minimum in this regime. After integrating \xco in each direction and averaging over the map, we obtain a median value for the CMZ component of \mbox{$ \rm{\overline{X}_{CO}^{CMZ}} = 0.39 \, \times$ \xcounit}. This median value is compatible with most of the results found in the literature that usually report average values for the CMZ in the range \mbox{$(0.2$ -- $0.7) \, \times$ \xcounit} \cite[see Table 3 of][and references therein]{2024PASJ...76..579K}. 

For the disk component, the distribution is also asymmetric with more low values but less peaked. The spread is larger with values in the range \mbox{$(0.49$ -- $2.87) \, \times$ \xcounit}, and the median is \mbox{$0.69 \, \times$ \xcounit}. Note that this median value for the disk is not comparable to the usual Galactic average value, first because it is only derived for a small region toward the GC, and secondly because of the approximations made in the \xco evolution model of this component discussed in the previous section.

A recent study from \citet{2024PASJ...76..579K} derived \xco maps in the CMZ from \co isotopologues using a very different method. They derived the \nhd from \twcoi lines using a constant \xco obtained for a given \xco(R) gradient as a function of the Galactocentric radius and the \nhd from \thcoi line using LTE approximation. Then they fitted an empirical curve to model the relation between the two estimates and used this to derive the \xco values. The drawbacks of this approach are: (i) it depends strongly on the choice of the reference \xco and metallicity gradient, (ii) the validity of the LTE approximation is not guaranteed in every direction for the extreme environment of the CMZ \citep[for details on the limitations of the LTE column density measurement of \thco see][]{2016MNRAS.460...82S}. So the \xco variations they report might also reflect the limit of these hypotheses. The \xco map they obtained is shown in Figure 3 of their paper. The comparison with our results is limited by the differences in resolution (ours is 10 times coarser), coverage (they show results only where \wco > 500 \wcounit), and CMZ component separation method.
However, the range of \xco values obtained is comparable, and the average estimates for the whole region are compatible. This agreement can be explained because the average value of \xco is mainly driven by the metallicity gradient assumed. Indeed they rely on the results from \citet{1996PASJ...48..275A} that yield a \xco $\propto \; Z^{-1}$, which is not far from the trend of $Z^{-0.8}$ we use as given by the simulation of \citet{2020ApJ...903..142G}.

Note that observations of CII and dust in external Galaxies suggest a trend in $Z^{-\alpha}$ with $\alpha \sim 1-2$ \textit{i.e} larger than the prescription from simulations we used, but the sample studied include mostly low-metallicity systems where steeper variations are expected \cite[see Figure 1b and 2a of][respectively, and references therein]{2024ApJ...961...42T, 2024ARA&A..62..369S}. Moreover comparison of the dependence in metallicity only is not necessarily straightforward as many other parameters play a role.

\begin{figure}
    \centering
    \includegraphics[width=\columnwidth, trim={0.5cm 0.5cm 0.5cm 0.5cm}, clip]{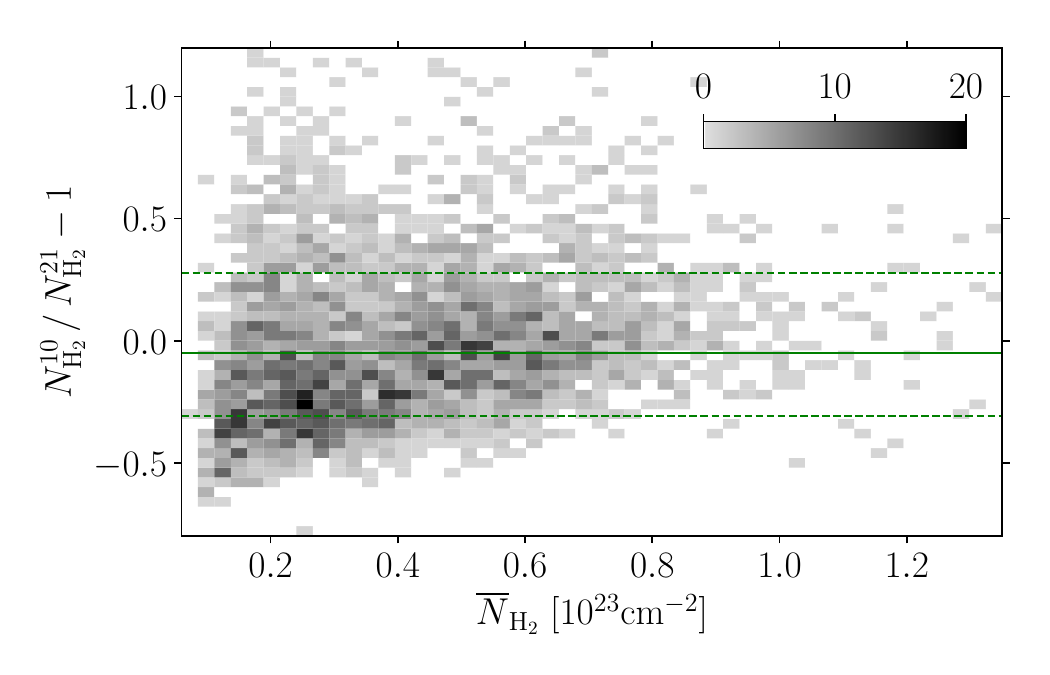}
    \caption{Relative uncertainty on the hydrogen column density in the molecular phase estimated from two different \co transitions as a function of the average estimate.
    Green solid line corresponds to the median ratio, dashed lines gives $1\sigma$ uncertainty  from the 16 and 84 percentiles.}
    \label{fig:nh2_errel}
\end{figure}

\begin{figure*}
    \centering
    \includegraphics[width=.9\textwidth, trim={8.9cm 3.1cm 7.2cm 4.3cm},clip]{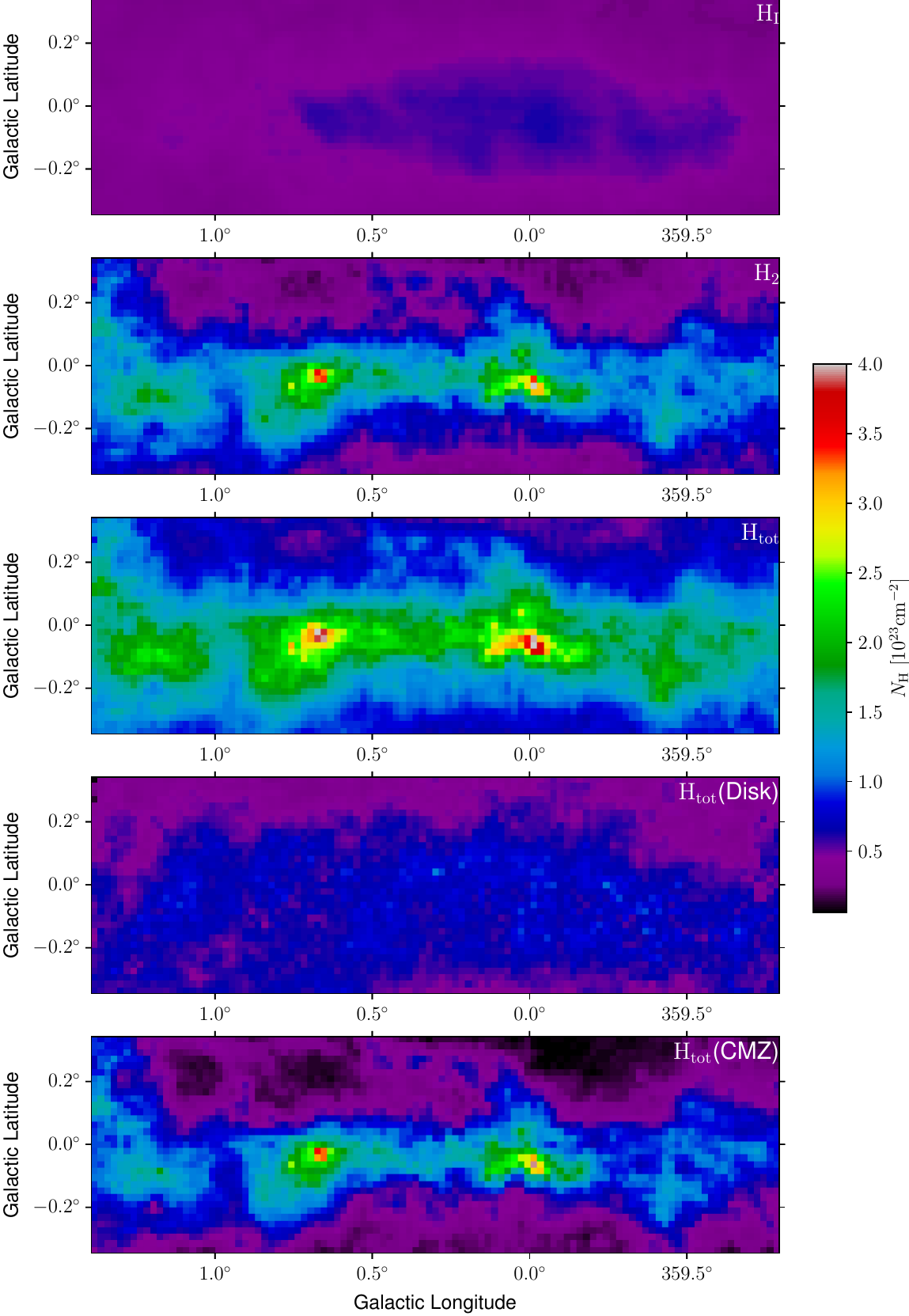}
    \caption{Hydrogen column density for the different phases \hi, \hd, $\rm H_{tot}=$ \hi + \hd, and the two components, disk and CMZ, separated. }
    \label{fig:nh_maps}
\end{figure*}

\begin{figure*}
    \centering
    \includegraphics[width=.9\textwidth, trim={0.95cm 1.3cm 0.5cm 0.3cm},clip]{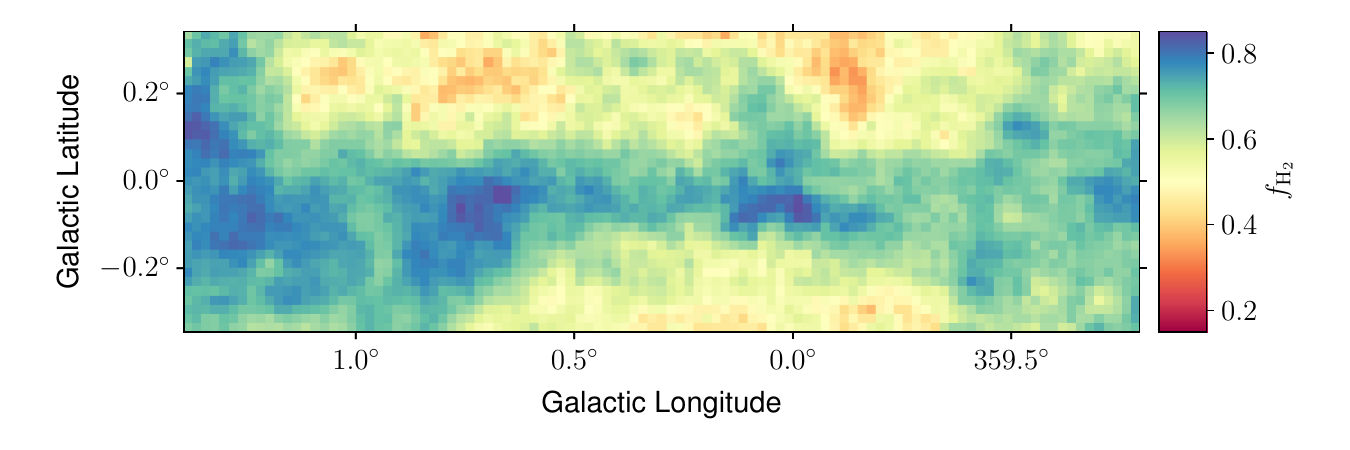}
    \includegraphics[width=.9\textwidth, trim={0.95cm 0.5cm 0.5cm 0.3cm},clip]{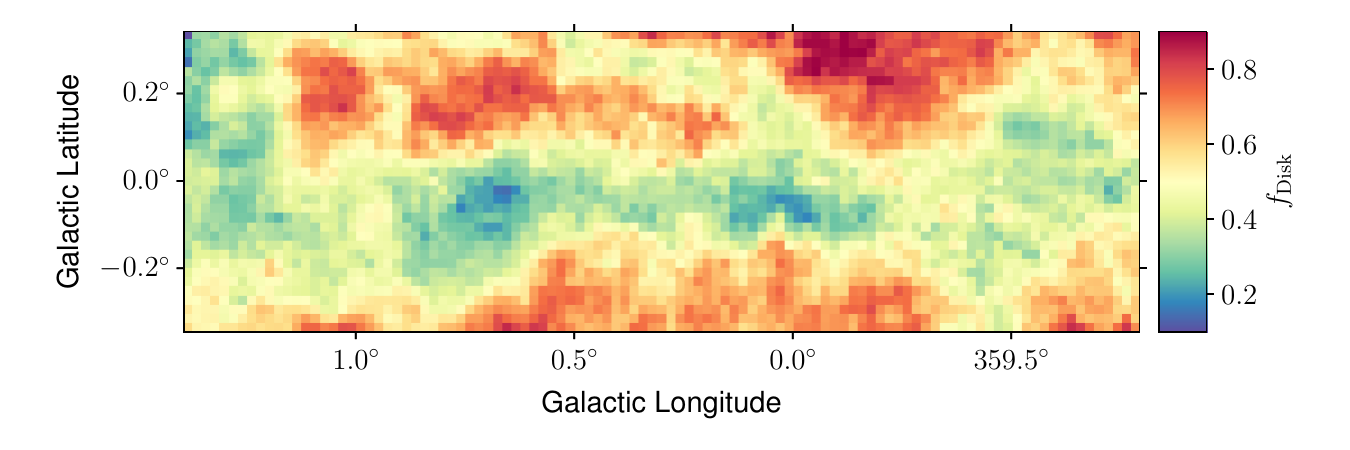}
    \caption{Top: Fraction of \hd in the total hydrogen column density. Bottom: Fraction of disk component in the total hydrogen column density.}
    \label{fig:fraction_maps}
\end{figure*}

\section{Hydrogen column density and masses per phase and per component}

\subsection{\nh maps products}\label{sec:res_maps}

The quantity of gas in the molecular phase is taken as the average of the two \nhd estimates obtained from Eq.~\ref{eq:nh2_12co} and \ref{eq:nh2_12co21}, which use the \xco factors derived for each fitted line. The relative uncertainty between the two estimates as a function of the mean is shown in Fig.~\ref{fig:nh2_errel}. 
On average, the $(2-1)$ estimate is 5\% higher, and with a 30\% dispersion. However, the uncertainty increases at lower \nhd and becomes more asymmetric. Indeed, at low \nh, the shape of the \xco(\wco) curves from the simulation is not corrected by the fit, and it cannot fully capture the variations of the \xco of the two tracers close to their sensitivity thresholds where the relative uncertainty on \wco is also large. In that regime, the use of higher sensitivity surveys and other tracers that can probe lower column densities could be considered to re-fit the \xco curves. However, such studies would be more suited toward isolated local clouds, for which we can better resolve the envelop of the clouds.
In the GC, this uncertainty on the evolution of \xco at low \nhd is not the main source of uncertainties as the absolute uncertainty is still dominated by higher column densities at \nhd $> 4.9 \times 10^{22}$ cm${^{-2}}$, and the molecular gas is not predominant at low column densities.

The hydrogen column density maps in atomic and molecular phases, as well as total maps for all the gas, the disk component and the CMZ component are shown in Fig.~\ref{fig:nh_maps}. The relative uncertainty with respect to the average \nhd estimate is shown in \ref{fig:err_nh2_map} in the appendix.
Maps of the fraction of molecular hydrogen, $f_{H_2}$, and of the fraction of the disk contamination, $f_{disk}$, in the total column density are displayed in Fig.~\ref{fig:fraction_maps}. The contribution of the atomic gas toward the GC is still $\sim$15-30\%; if we remove the disk contamination, this value is reduced to $\sim$5-15\% for the CMZ clouds. The atomic gas fraction increases to  
50-70\% in the regions that are dominated by the disk component at only 0.2$\degree$ from the CMZ clouds. This is particularly important for \g-ray analyses as the diffuse Galactic emission comes mostly from the \g rays produced by the interaction of CR with all the interstellar gas. Then, the contribution of the atomic phase and the disk gas contamination should not be neglected, even toward the CMZ clouds.

\subsection{Mass estimates}\label{sec:mass} 

The gas mass, $M$, is given as a function of the total hydrogen column density, \nh $= N_{\rm HI} + N_{\rm H_2}$, by :
\begin{equation}
   M = \mu\, m_{\rm H} \, d^2 \sum_{\rm pixels} N_{\rm H} \, \rm d \Omega,
\end{equation}
with $d$ the distance to the observer, $\rm d \Omega$ the solid angle, \mh the atomic mass, and $\mu = 1.4 $ the mean molecular weight \citep[same as in][]{2022MNRAS.516..907S}.

The resulting gas masses in the atomic and molecular phases and the total value toward the GC within $-0.8\degree< l <1.4\degree$ and $|b|<0.3\degree$ are presented in Table \ref{tab:mass}. The table also provides the mass for CMZ and disk components separately. Note that the value for the disk is only computed considering the distance of the CMZ in order to evaluate the bias on the total mass estimate that would be introduced by the absence of component separation (as typically given by dust-based estimates).
The total gas mass in the CMZ is found to be $2.3_{-0.3}^{+0.3}\times10^{7} \; \rm{M_{\odot}}$ with only $\sim$10\% contribution from the atomic gas. If we also consider the gas in the disk component, we reach $4.4_{-0.7}^{+0.3}\times10^{7} \; \rm{M_{\odot}}$ and the atomic gas contribution rises to $\sim$30\%.

\renewcommand{\arraystretch}{1.8}
\setlength{\tabcolsep}{0.3cm}
\begin{table}
    \caption{Gas mass per phase and per component in solar masses toward the GC within $-0.8\degree< l <1.4\degree$ and $|b|<0.3\degree$.}
\centering
    \begin{tabular}{ c | c c c }\hline \hline  & \hi$^{(a)}$ & $\rm H_2^{(b)}$ & $\rm{H_{tot}}$ \\ \hline Disk$^{(c)}$ &  $1.2_{-0.3}^{+0.0}\times10^{7}$ & $8.9_{-1.8}^{+1.8}\times10^{6}$ & $2.1_{-0.5}^{+0.2}\times10^{7}$ \\ CMZ \; & $2.7_{-0.6}^{+0.0}\times10^{6}$ & $2.0_{-0.3}^{+0.3}\times10^{7}$ & $2.3_{-0.3}^{+0.3}\times10^{7}$ \\ Total$^{(c)}$ & $1.4_{-0.3}^{+0.0}\times10^{7}$  & $2.9_{-0.3}^{+0.3}\times10^{7}$ & $4.4_{-0.7}^{+0.3}\times10^{7}$  \\\hline \hline \end{tabular}

\label{tab:mass}
\tablefoot{
    \tablefoottext{a} {The lower uncertainty corresponds to a milder \hi absorption correction.}
    \tablefoottext{b} {Positive/negative errors correspond to the min/max deviation from the average \nhd estimate between the two \co isotopes considered.}
    \tablefoottext{c} {The mass in the disk is only computed considering the distance of the CMZ in order to evaluate the bias on the total mass estimate that would be introduced by the absence of component separation.}
    }
\end{table}
    
\begin{figure*}
    \centering
        \includegraphics[width=.9\textwidth, trim={1.1cm 1.3cm 0.15cm 0.3cm},clip]{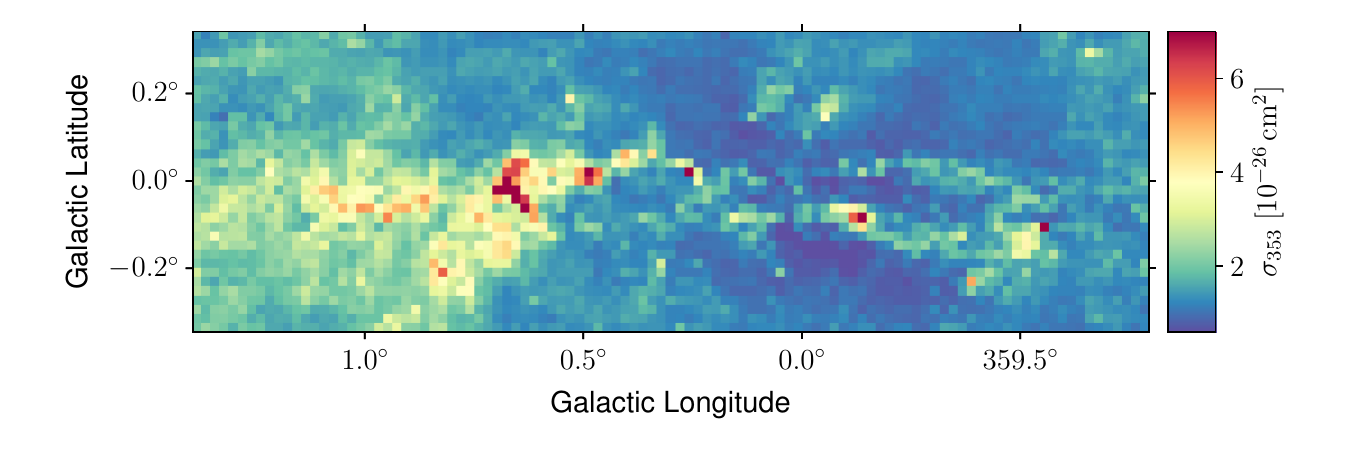}
        \includegraphics[width=.9\textwidth, trim={0.90cm 0.cm 0.15cm 0.3cm},clip]{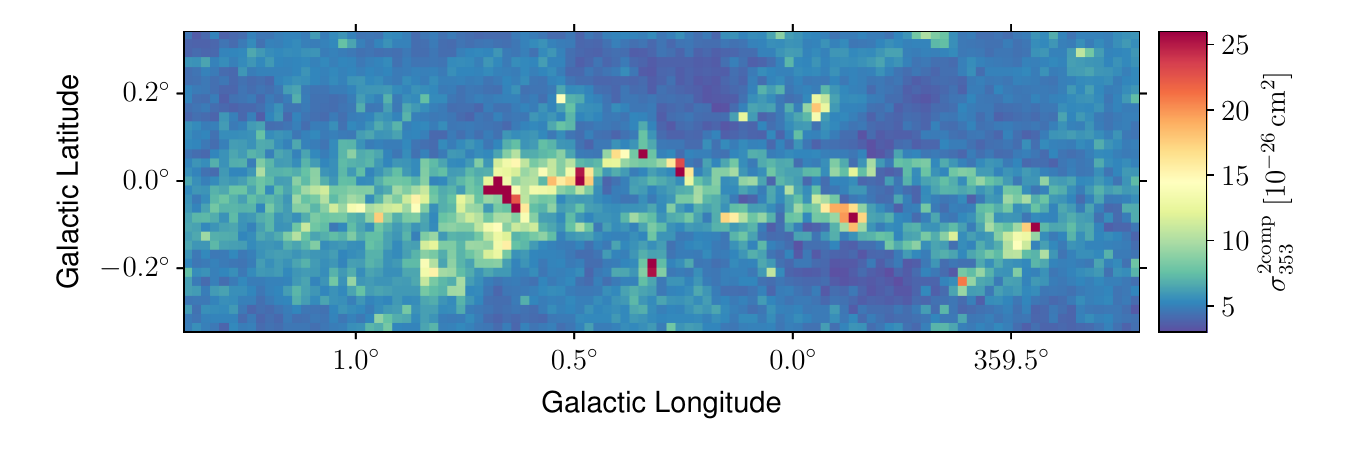}
    \caption{
    Top: Dust opacity at 353 GHz, $\sigma_{353} =$ \opa, derived from the single component fit.
    Bottom: Total dust opacity at 353 GHz, derived from the two-components fit.
    In both cases \nh is the total hydrogen column density shown in the third panel of Fig.~\ref{fig:nh_maps}.
}
    \label{fig:tau_opa}
\end{figure*}

The masses derived in the molecular phase are very similar to those of \citet{2022MNRAS.516..907S} that found values $2.8\times 10^7  \; \rm{M_{\odot}}$ in total and $2.3\times 10^7  \; \rm{M_{\odot}}$ in the CMZ after removing the disk contamination (with another method). The small differences could be due to their slightly different region definition ($-1.1\degree< l <1.8\degree$ and $|b|<0.2\degree$) and the use of a constant \xco factor.
In the atomic phase, we found a total mass 3 times higher than the values they reported. This discrepancy can be simply explained by two factors: (i) they derived a minimal mass for the optically thin case while we use a spin temperature of $T_{\rm S}=146.2\pm16.1~\mathrm{K}$ \citep{2017MNRAS.468.4030S} which yield to 1.6 times higher mass; (ii) the method we use to correct for self-absorption allows to recover about 1.7 times more mass.
Even with this higher \hi mass, the molecular gas fraction we find in the CMZ clouds after removing the disk contamination is still as high as $\sim$85-95\% and decreases only by a few per cent compared to the values of $\sim$90-98\% reported by \citet{2022MNRAS.516.3911S}.

\begin{figure}
    \centering
    \includegraphics[width=\columnwidth, trim={0.5cm 0cm 0.5cm 0.cm}, clip]{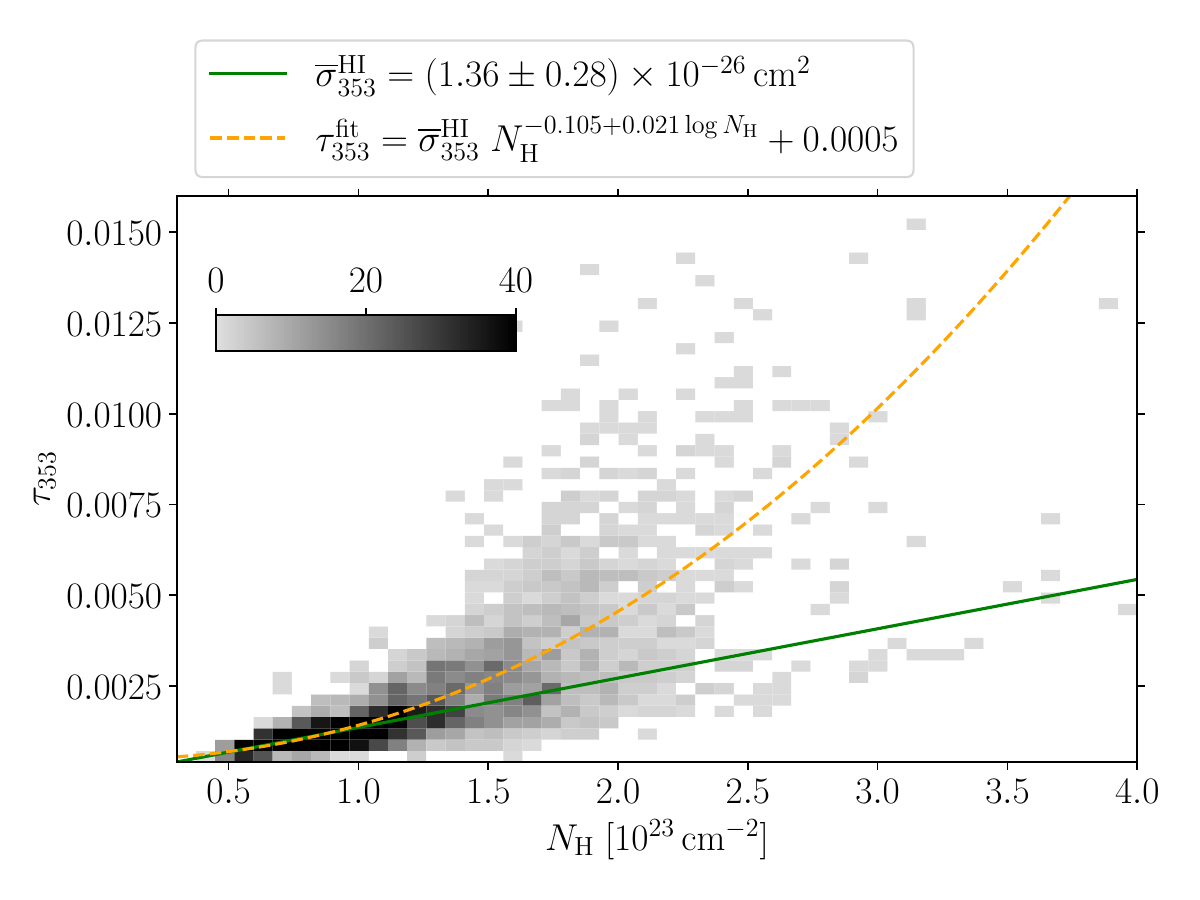}
     \includegraphics[width=\columnwidth, trim={1.4cm 0cm 1.4cm 0cm},clip]{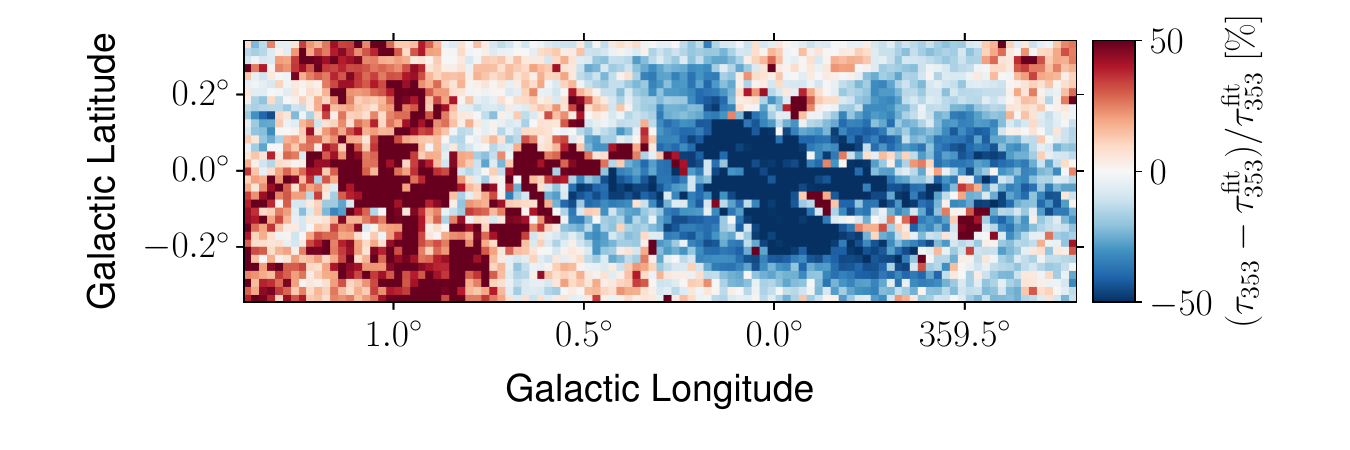}

    \caption{Top: Dust optical depth as a function of \nh.
 			The green line corresponds to a linear relation for which the slope is the mean opacity in the regions dominated by atomic gas ($f_{H_2}<0.5$). The orange curve corresponds to the best-fit \taunu as a function of \nh with a log-parabola model. Bottom: Relative error on the fit.}
    \label{fig:nh_opa}
\end{figure}

\section{Dust opacity variations}\label{sec:dust_discussion} 
The optical depth, $\tau_\nu$,  can be related to the dust mass, $M_{\rm dust}$, and gas column density, \nh, as follow :
\begin{align}
    \tau_\nu &= \kappa_\nu M_{\rm dust} \\
    \tau_\nu &= \sigma_\nu N_{\rm H},
    \label{eq:opa}
\end{align}
where $\kappa_\nu$ is the dust emissivity cross section per unit mass, \mbox{$\sigma_\nu = \kappa_\nu r_{\rm dg} \mu $\mh} is the dust opacity, $r_{\rm dg}$ is the dust-to-gas mass ratio, and $\mu$ is the mean molecular weight. 

The opacities derived at 353 GHz from the optical depth obtained with the single, and two-component dust models are shown in the top, and bottom panel of Fig.~\ref{fig:tau_opa}, respectively. The values derived for the two-component model are systematically higher by a factor $2$ to $12$. We also note differences comparing our two-components model to one of \citet{2015ApJ...798...88M}, which has less free parameters (we fit $\beta_{\rm cold}$, $\beta_{\rm warm}$ and $T_{\rm cold}$ in each line of sight). The values of the optical depth at 545 GHz, $\tau_{545}$, we obtain at the same resolution are within a factor $0.85-2.3$ compared to the $\tau_{545}$ values derived by \citet{2015ApJ...798...88M}. 
This illustrates that the hypotheses used for the dust emission fit have a strong impact on the results obtained.

\begin{figure*}
    \centering
    \includegraphics[width=\textwidth, trim={0.5cm 0.5cm 0.5cm 0.5cm}, clip]{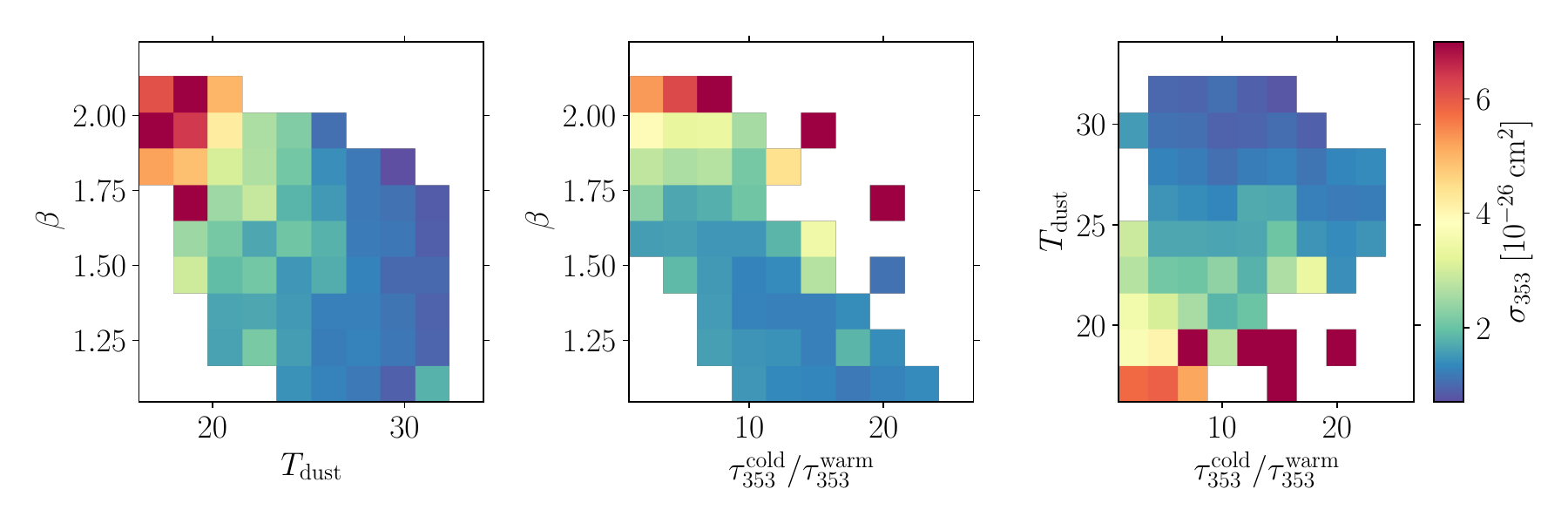}
    \caption{Evolution of the dust opacity with dust spectral index $\beta$, dust temperature  $T_{\rm dust}$ (single-component), and the cold to warm components ratio in dust optical depth, \tauratio.}
    \label{fig:opa_grid}
\end{figure*}

In the following we discuss the dust optical depth and opacity values derived from the single-component dust model as there are more references for comparisons in the literature.
The average opacity in the \hi dominated regions is $1.36 \pm 0.28$~\opaunit. Similarly regions dominated by the disk component (which mostly correspond to regions dominated by atomic gas) have an average opacity of  $1.39 \pm 0.43$~\opaunit. These values are quite close to what is found in the \hi phase of local clouds (see Tables 2 of \citet{2015A&A...582A..31P} and \citet{2017A&A...601A..78R}) and twice larger than the values found in the most diffuse regions at high latitudes \citep[see table 4 of][]{2014A&A...571A..11P}. Large variations up to a factor of 9 above these averages are found toward dense molecular clouds such as Sgr~B2 and Sgr~C. This trend is again similar to what is found in local clouds but in the GC more extreme values are reached.

In Fig.~\ref{fig:nh_opa} we attempt to model the variation of \taunu as a function of \nh with a non-linear relation as done in earlier studies \citep{2013ApJ...763...55R, 2017ApJ...838..132O, 2019ApJ...878..131H}. However we can see that there is a large dispersion around the fitted trend, and the fit residuals are very structured. This suggests that assuming only the non linearity of \nh is not sufficient to model the dust opacity variations and that other mechanisms are at play.

The variations of the opacity with dust spectral index, temperature, and cold-to-warm optical depth ratio are shown in Fig.~\ref{fig:opa_grid}. 
In the left panel we show that the opacity raises with increasing $\beta$ and decreasing $T_{\rm dust}$ as observed in local clouds \citep[see Figure 18 of][]{2017A&A...601A..78R}.
Theoretical works suggest that grains evolution in denser media involving consecutively the accretion of carbonaceous mantles, grain coagulation, and ice mantle formation progressively alter the emissivity of the grains. These processes can explain the increase in opacity with gas density (up to a factor 7) and the associated changes in spectral index $\beta$, and colour temperature as shown by \citet{2015A&A...579A..15K}.

\begin{figure}
    \centering
    \includegraphics[width=\columnwidth, trim={0.5cm 0.5cm 0.5cm 0.5cm}, clip]{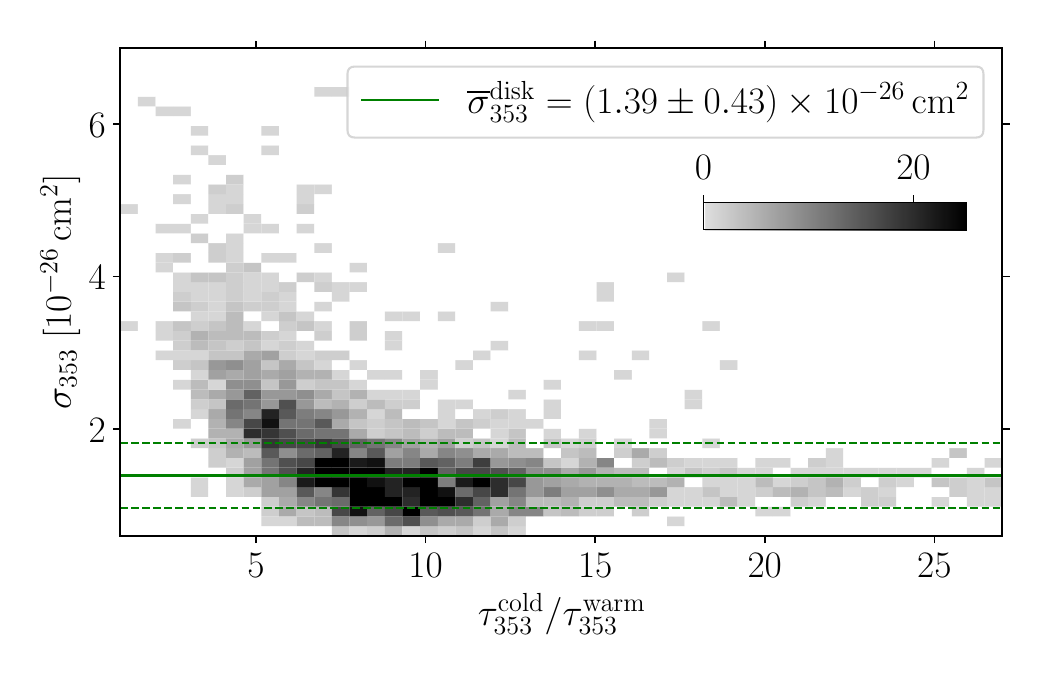}
    
    \caption{Dust opacity as a function of the cold to warm components ratio in dust optical depth, \tauratio.
    The green line corresponds to the mean opacity in the regions dominated by the disk gas ($f_{\rm disk} >0.5$) which mostly correspond to regions dominated by atomic gas.}
    \label{fig:tau_ratio_opa}
\end{figure}

Despite the agreement with the trend of variations seen in local clouds, the ranges of opacity, spectral index and colour temperature values measured in the CMZ are broader.
In Fig.~\ref{fig:tau_ratio_opa} we can see that the opacity increases as the cold-to-warm dust ratio in optical depth decreases. The enhanced contamination from warm dust, interestingly correlates not only with an increase in opacity but also with an increase in $\beta$, and a decrease in colour temperature (see middle and right panels of Fig.~\ref{fig:opa_grid}, respectively). It is unclear whether this effect is due to a degeneracy of the parameters fitted to the spectral energy distribution, or to a physical process. As pointed by \citet{2019A&A...631A..88Y} the spectral index measured on dust analogues in laboratory are not always straightforward to compare with the ones derived from astronomical observations. Indeed the observed spectral index can be significantly altered by the dust temperature distribution along the line of sight, which is influenced by heating sources and the structure of the ISM.

Other possible causes of increase in opacity are the raise of the dust-to-gas mass ratio with enhanced metallicity, and turbulence.
The correlation between dust-to-gas mass ratio and metallicity reflects the progressive chemical enrichment of the ISM with dust created from heavy elements ejected by stars at the end of their lifetime and by stellar winds. However the correlation is more marked at low metallicity, while at solar metallicity and above the dust-to-gas mass ratio seems to flatten \cite[see Figure 1, 3, and 8 of ][respectively, and references therein]{2011A&A...532A..56G, 2019MNRAS.490.1425L, 2021A&A...649A..18G}.
The enhanced turbulence level in the CMZ could also increase the dust-to-gas mass ratio, indeed \citet{2017MNRAS.471L..52T} found evidence of grain-size dependent sorting of the dust, because turbulence preferentially concentrates larger grains into dense regions.

Given the complex dependency of the changes in dust opacity with the unknown dust composition and the environment all along the line of sight toward the GC, it is very complicated to rely on a physical model of dust evolution. Moreover the empirical models describing the dust opacity are hardly transposable from one region to another. In particular gas mass estimated from the dust emission using a simple averaged opacity as scaling factor would be highly unreliable. For example considering the average opacity of $8.4 \times 10^{-27}$ cm$^{2}$ measured by \citet{2014A&A...571A..11P} over the whole sky would yield to a CMZ mass of $1\times 10^{8} \; \rm{M_\odot}$, more than a factor 2 higher compared to the total mass estimation discussed in Sect. \ref{sec:mass} \citep[and also higher than historical measurements reported in Table 2 of][]{1998A&A...331..959D}. If we take into account the dust evolution in different media with a minimal model considering an average opacity of $14 \times 10^{-27}$ cm$^{2}$ in the \hi dominated regions and an opacity enhanced by a factor $1.5$--$3$ on average for \hd dominated regions (similar to what is found in local clouds), it yields masses of $2.2$--$4.1\times 10^{7} \; \rm{M_\odot}$ comparable to the total gas estimates from \hi $+$ CO lines. Toward the CMZ, and the Galactic plane in general, dust is not necessarily a more reliable tracer of the total gas because the opacity evolution with environment is as complex, if not more, than the evolution of the \xco factor.

\section{Impact on cosmic-ray density estimation}\label{sec:CR}

The diffuse \g-ray Galactic emission comes mostly from the \g rays produced by the CR interaction with the interstellar gas (as a product of the decay of neutral pions created in proton-proton interactions).
Assuming that other contributions to the diffuse \g-ray emission are negligible, the CR energy density can be expressed as a function of the \g-ray luminosity $L_\gamma$ above a given energy $E_\gamma$, and the gas mass $M$ by:
\begin{equation}
    w_{\rm CR}(\geq 10 E_\upgamma) \simeq 0.018\left(\frac{1.5}{\eta_{\rm N}}\right) \left(\frac{L_{\upgamma}(\geq E_\upgamma) }{10^{34} \, \rm{erg/s}}\right) \left(\frac{10^6 \, \rm{M_\odot}}{M}\right) \rm{eV/cm^3}
    \label{eq:wcr}
\end{equation}
where $\eta_{\rm N} = 1.5$ accounts for the presence of nuclei heavier than hydrogen in both CR and interstellar matter.

\citet{2016Natur.531..476H} used this relation\footnote{\label{fn:HESS} Note that there is an inconsistency in \citet{2016Natur.531..476H}: the $ w_{\rm CR}$ values given in its Table 2 do not match the values obtained from this equation when we recompute them from the data in its Table 1. We suspect that the  $ w_{\rm CR}$ values reported in the H.E.S.S. paper were wrongly computed with a value of $1.4 \times 10^8 \, \rm{yrs \, (cm^{-3}}/N_{\rm H})$ for the proton energy loss timescale $t_{pp\rightarrow\pi^0}$ instead of the value of $1.6 \times 10^8 \, \rm{yrs \, (cm^{-3}}/N_{\rm H})$ quoted in the paper.} to derive the CR density above 10 TeV ($E_\gamma \geq 1$ TeV ) in several regions toward the CMZ (see Figure 1 and 2 of their paper). In their work the mass in each region is simply taken as a fraction of the total mass, assumed to be $3 \times 10^7 \, \rm M_{\odot}$, weighted by the integrated intensity in the region compared to the whole map. Three different estimates are given based on different tracers: CS, \co and HCN lines. This approach implicitly assumes a constant conversion factor, ignores the gas below the critical density of the tracers considered, and ignores the contribution of the atomic gas which increases with increasing absolute latitudes. As mentioned previously, CRs interact with the interstellar gas in all its phases, so the contribution from both the diffuse atomic hydrogen and the lower-density molecular gas in the disk should not be neglected, even toward the CMZ clouds.

\begin{figure}
    \centering
    \includegraphics[width=0.99\columnwidth, trim={0.35cm 0.7cm 0.3cm 0.6cm},clip]{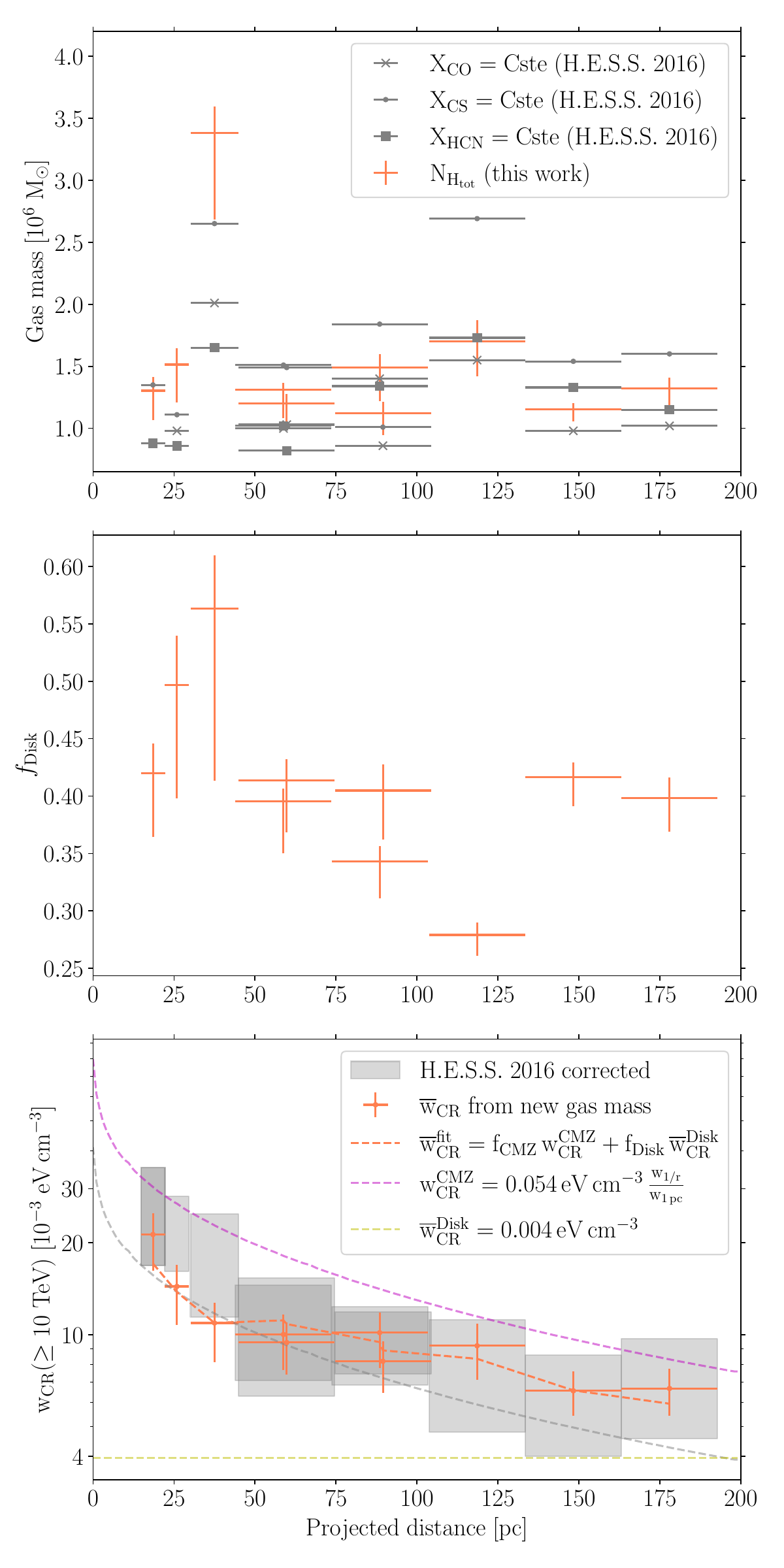}
    \caption{
    Top: Gas mass estimates from \citet{2016Natur.531..476H} in grey and from this work in orange.
    Middle: Fraction of disk contamination in the total hydrogen column density.
    Bottom: Average CR density as a function of its projected distance from Sgr~A*. The grey items are adapted from \citet{2016Natur.531..476H}, including the systematic uncertainties associated to the choice of gas tracer, and the correction we discuss in the footnote \ref{fn:HESS}. The grey line is the 1/r profile in CR density integrated over the line of sight provided in their paper as a fit to the data points. The orange error bars give the values derived from our mass estimates. The orange line corresponds to the fit to these points considering a 1/r profile for the CMZ component (magenta) and a constant for the disk component (yellow).
}
    \label{fig:mass_cr}
\end{figure}

We recompute the CR energy densities in the regions they analysed considering our total gas mass estimate. The results are summarized in Fig.~\ref{fig:mass_cr}; the top panel shows the gas mass comparison, the middle panel shows the fraction of disk contamination in the total hydrogen column density, and the bottom panel shows the CR energy density. The largest deviation in mass and CR density is found in the three first regions that include more disk contribution because they extend at higher latitudes. 

\citet{2016Natur.531..476H} fitted the $w_{\rm CR}$ values with a 1/r profile in CR density integrated over the line of sight which suggests a continuous injection from a central accelerator (potentially Sgr~A$^\star$, the central super massive black hole). However, the CR density derived from the total \g-ray luminosity and total gas mass is a weighted average over the line of sight, so their analyses implicitly assume that most of the \g rays originate from the CMZ and that contribution from the disk is negligible. This assumption may not hold because: (i) We shown here that the gas mass in the disk contributes to 25-60\% in the regions considered; (ii) Several studies reported an enhancement of the CR density (or \g-ray emissivity per gas nucleon) in the disk up to a factor of three compared to the local value (see Figure 3 at the end of \citet{2021Univ....7..141T} for a compilation of Fermi-LAT at GeV energies, and Figure 2 of \citet{2023PhRvL.131o1001C} for LHAASO measurement at TeV energies). \citet{2016Natur.531..476H} calculated a local value of \mbox{$ w_{\rm CR, \, local}(\geq 10 \, \rm TeV) \sim 10 ^{-3} \; \rm{eV \,  cm^{-3}}$}, so an enhancement factor of three in the disk would already account for half of the CR density found in two of the regions considered, therefore it is unlikely to be negligible.

In order to take into consideration the contamination from the disk, we fitted the $\rm{{w}_{CR}}$ values with two components via a \chisq minimization :
\begin{equation}
    \overline{w}_{\rm{CR}}^{\rm fit} = f_{\rm CMZ} \, w_{\rm{CR}}^{CMZ} + f_{\rm Disk} \, \overline{w}_{\rm{CR}}^{Disk}.
\end{equation}
The fractions $f_{\rm CMZ}$ and $f_{\rm Disk}$ are the weight of the CMZ and disk components in the total gas column density, respectively. The CR energy density in the CMZ is modelled with the same 1/r profile integrated over the line of sight as in \citet{2016Natur.531..476H}, noted $w_{1/r}$, for which we refit only a normalisation, and find \mbox{$ w_{\rm{CR}}^{\rm{CMZ}} = \left(53.9\pm 10.7 \right) \times 10 ^{-3} \, w_{1/r}/w_{1\, \rm pc} \; \rm{eV \,  cm^{-3}} $}. The other fitted parameter is \mbox{$\overline{w}_{\rm{CR}}^{\rm{Disk}} = \left(4.0 \pm 3.6 \right) \times 10 ^{-3} \; \rm{eV \,  cm^{-3}}$} that gives the average level of the CR sea in the disk toward the GC. Even if the uncertainties are large, the two-component model clearly improves the fit by 4.9 $\sigma$ over a single uniform value and by 2.6 $\sigma$ over a 1/r profile only. Considering two components yields to about two times larger  $w_{\rm CR}$ in the CMZ compared to the previous study, since our result is not biased downward by the mixture with disk contribution that has lower $w_{\rm CR}$ values. Interestingly the average value found in the disk is a factor of four times the local value, which is consistent with the enhancement reported in the studies cited in the previous paragraph. However, the uncertainty is large, so considering more regions and larger regions would be required to further discuss the CR densities across the Galactic disk.
Note also that as in \citet{2016Natur.531..476H}, this fit does not take into account the 3D distribution of the gas across the CMZ, which limits the discussion on the validity of the 1/r distribution for the CR densities. We defer more refined studies considering the distance of the different clouds in the CMZ to future works.

\section{Summary and perspectives}\label{sec:ccl}

In this study, we analysed the GC region defined by longitudes $-0.8\degree< l <1.4\degree$ and latitudes $<0.3\degree$, at an angular resolution of $\sim 0.02 \degree$.
We have calculated the total hydrogen column density, taking into account both the atomic and molecular components of the gas, and the contribution from both the CMZ and the foreground/background gas in the disk.
 
The atomic phase was probed using the \hi Galactic Center survey data from ATCA and the Parkes Radio Telescope \citep{2012ApJS..199...12M}. A \hi self-absorption correction was applied, to recover more atomic gas in the central $\sim1 \degree$ radius around the GC. Then, the atomic hydrogen column density has been computed with a spin temperature  $T_{\rm S} \simeq 150$ K \citep{2017MNRAS.468.4030S}. 

For the molecular phase, \co isotopologue line emissions has been used: \twco and \thco for the $J=1 \rightarrow 0$ transition from the NRO \citep{2019PASJ...71S..19T}, and \thco and \eico for the $J=2 \rightarrow 1$ transition from the APEX telescope \citep{2021MNRAS.500.3064S}. A baseline correction was applied to the molecular line data cubes to improve sensitivity and reduce discrepancies across the map. Isotope ratio in integrated intensities have been found to be: $ \overline{R}_{1813} = 0.10 ^{+0.04} _{-0.03}$ for the median ratio of \weicoii over \wthcoii and $ \overline{R}_{1312} = 0.14 ^{+0.04} _{-0.03}$ for the median ratio of \wthcoi over \wtwcoi, which is compatible with the value published by \citet{2019PASJ...71S..19T}.

For each tracer considered and for each (l,b) directions the brightness temperature profiles as a function of velocity were decomposed in multiple line contributions. This profile decomposition helps to identify changes in gas properties. 
Indeed in the CMZ, higher turbulence and shear lead to greater velocity dispersion within and between clouds, causing emission lines to be more spread apart in velocity compared to the Galactic disk. This increased turbulence also decreases the optical thickness of the \twco emission lines, raising their maximal brightness temperatures, resulting in a higher ratio of brightness temperatures  $T_{^{12} \rm{CO} }/T_{^{13}\rm{CO}}$ in the CMZ than in the Galactic disk. These differences in properties were used to perform a component separation between the gas emission from the CMZ and from the Galactic disk.
The profile decomposition and the component separation allowed us to derive a \xco value for each line considering different \xco evolution models for the CMZ and the disk gas.

The evolution of the \xco factor as a function of CO integrated intensity, metallicity, and resolved scaled, was modelled using theoretical trends from simulations in \citet{2020ApJ...903..142G}. An empirical power-law correction was added to take into account the effect of the increased level of turbulence in the CMZ following \citet{2016MNRAS.455.3763B}. The index of the power-law correction $\eta$ for each component were derived using a fit by minimizing the difference between the gas estimates found adopting different isotopes. In the main analysis the average metallicity in the CMZ is fixed to a value from the literature, but if left as a free parameter we found a value $\sim +0.3$ dex which is compatible with the super-solar metallicities reported in several studies \citep{2015ApJ...809..143D,  2018MNRAS.478.4374N, 2019A&A...627A.152S}.
For the CMZ component, the \xco values per line range from \mbox{$(0.32$ -- $1.37) \times$ \xcounit}, with a distribution that is highly asymmetric and skewed toward the minimum value. This is because most detected lines have \wco values between a few tens and a few hundreds of \wcounit, where the \xco(\wco) profile is nearly flat. After integrating \xco in each direction and averaging over the map, the median value for the CMZ component is found to be \mbox{$ \rm{\overline{X}_{CO}^{CMZ}} = 0.39 \times$ \xcounit}.

Separate maps for the atomic, molecular and total gas were computed for the CMZ and the Galactic disk components. The total gas mass estimated in the CMZ is $2.3_{-0.3}^{+0.3}\times10^{7} \; \rm{M_{\odot}}$ with only ~10\% contribution from the atomic gas \citep[similar to the estimates from][]{2022MNRAS.516..907S}. Without removing the disk contamination the total mass is about twice higher, and the atomic gas fraction increases to $\sim30\%$. Therefore the contamination from the gas in the disk is not negligible, and the same holds for the \hi contribution to the total gas mass in the GC region as well.
 
Modelling the dust opacity changes toward the GC (and the Galactic plane in general) is very complex due to the unknown dust composition and varying environments along the line of sight. The gas mass estimates from dust emission are far to be free from uncertainties, and we stress that the use of a simple average opacity as scaling factor would be very unreliable, and would lead to mass overestimation.

Estimating the total gas content in the GC region is particularly important for \g-ray astronomy, as it plays a key role in determining the CR density from the observed \g-ray emission and the total mass of interstellar gas.
We recalculated the CR energy density $\rm{{w}_{CR}}$ using the formalism adopted in \citet{2016Natur.531..476H}, and our new estimates of the total gas mass. Firstly, we found consistent gas mass for the different regions selected, except for their first three regions, where the disk contribution is higher due to their extent up to high latitudes. Secondly, we fitted the CR energy density accounting for two components, the CMZ and the Galactic disk, in order to take into account the contamination from foreground/background gas that was previously neglected. The CMZ component was modelled with the same 1/r profile, while the added disk component represents the average level of the CR in the disk toward the GC. We fond that the two-component model is preferred by 4.9$\sigma$ over a uniform value, and by 2.6$\sigma$ over the 1/r profile. As a result, the CR energy density in the CMZ is higher by a factor of two compared to previous measurements, and the average value for the disk toward the GC is found to be four times higher than the local value.

To further refine our model, it’s necessary to consider the three-dimensional distribution of CRs and gas in the GC. However, this aspect will be addressed in a subsequent study. Understanding the 3D gas distribution in the CMZ is crucial for comprehending the formation and evolution of this structure, as well as for addressing broader questions related to star formation sites and CR transport \citep{2017arXiv170505332M, dör24}.
There are already some indications about the structure of the CMZ. For instance, \citet{fer07} suggested that it has an asymmetric ring-like shape. Additionally, \citet{sch22} deduced from \g-ray maps that an inner cavity must be present. The giant molecular clouds in the CMZ are thought to follow a ‘twisted-ring’ orbit, as proposed by \citet{mol11}. Nevertheless, the exact spatial configuration of these clouds on the orbit remains a topic of debate \citep{hen23}.
To achieve a more precise 3D mapping, the region should first be divided into individual clouds using clustering algorithms similar to those described by \citet{2017ApJ...834...57M}. The position of these clouds can then be determined using one of two methods: directly estimating the line of sight distance through absorption lines \citep{yan17, 2022MNRAS.516..907S}, or calculating the distance to the supermassive black hole (SMBH) by analysing the clouds’ response to the SMBH’s activity \citep{mar14, ter18}. This problematic of the 3D distribution of CRs and gas in the GC will be explored in future studies.

\begin{acknowledgements}

The authors thank S.~C.~O.~Glover and J.~A.~Hinton for providing their valuable comments and suggestions on the draft, and R.~J.~Tuffs for the interesting discussions. 
S.R. acknowledges support from the LabEx UnivEarthS, ANR-10-LABX-0023 and ANR-18-IDEX-0001. S.R. and M.B. acknowledge funding from SALTO exchange program between the CNRS and the MPG.
H.R. expresses her gratitude to the support from the International Max Planck Research School for Astronomy and Cosmic Physics at the University of Heidelberg (IMPRS-HD).  
J.D. acknowledges funding from the Research Council of Norway, project number 301718.
We thank the communities behind the multiple open-source software packages on which we depend, in particular astropy \citep{2022ApJ...935..167A} for FITS files and units handling, gammapy \citep{gammapy:2023} for skymaps utilities, matplotlib \citep{Hunter:2007} for visualisation, and numpy \citep{harris2020array}, scikit-image \citep{scikit-image}, scikit-learn \citep{scikit-learn}, scipy \citep{2020SciPy-NMeth}, lmfit \citep{matt_newville_2023_8145703} for all the computation methods they offer.

\end{acknowledgements}

\bibliographystyle{aa_url}
\bibliography{biblio}

\begin{appendix}
\onecolumn
\section{Additional figures and tables}

\begin{figure}[H]
    \centering
    \includegraphics[width=0.4\textwidth, trim={0.5cm 1.5cm 0.0cm 1.5cm},clip]{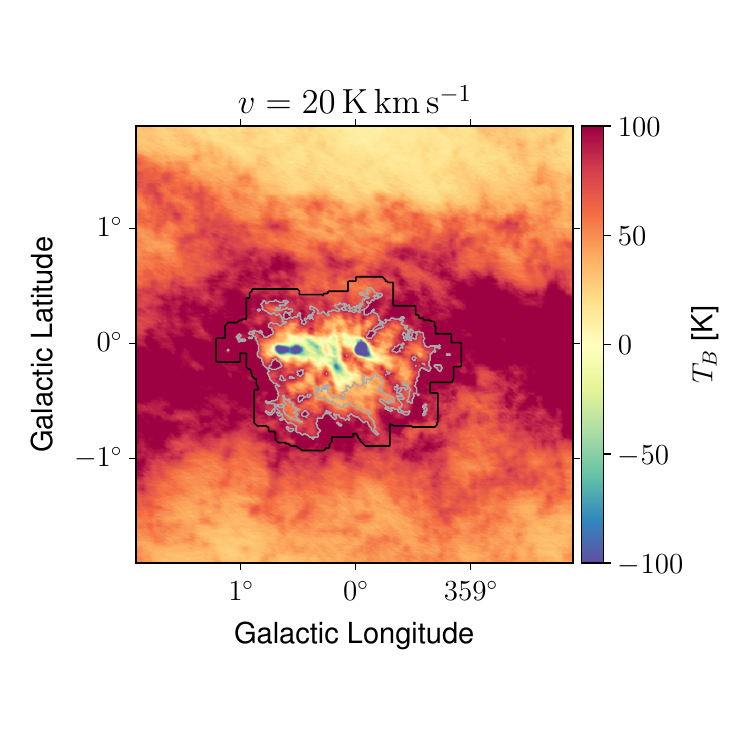}
    \includegraphics[width=0.55\textwidth, trim={0cm 0.cm 0.5cm 0.5cm},clip]{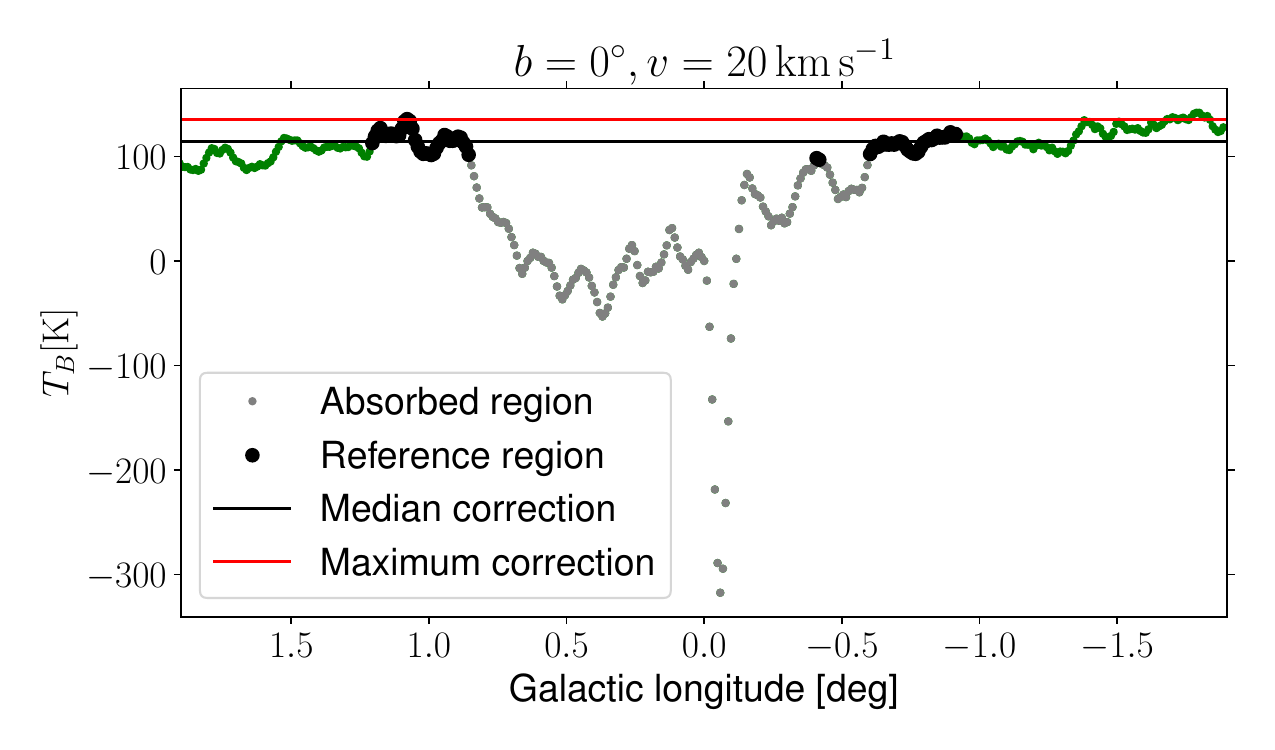}
    \caption{Left: \hi temperature brightness at $v=20 \, \rm{km} \, \rm{s}^{-1}$. Gray contours delimit the regions we consider affected by absorption  (because of low temperature brightness and large gradient values). Black contours correspond to a 10 pixels margin around the grey contours that we use as reference region to compute the median or maximum values in the vicinity. Right: \hi temperature brightness at $v=20 \, \rm{km} \, \rm{s}^{-1}$ and $b=0\degree$. The median and maximum values are computed along each longitude value considering only the reference pixels (black) and used as substitute for the absorbed pixels (grey). The same operation is applied for each velocity independently. More details are given is the text of Sect.~\ref{sec:hi_abs}, and the resulting \nh maps are shown in Fig.~\ref{fig:NHImaps}.
    }
    \label{fig:NHImethod}
\end{figure}

\begin{figure}[H]
    \centering
    \includegraphics[width=\textwidth, trim={2.8cm 18cm 1.5cm 2.9cm},clip]{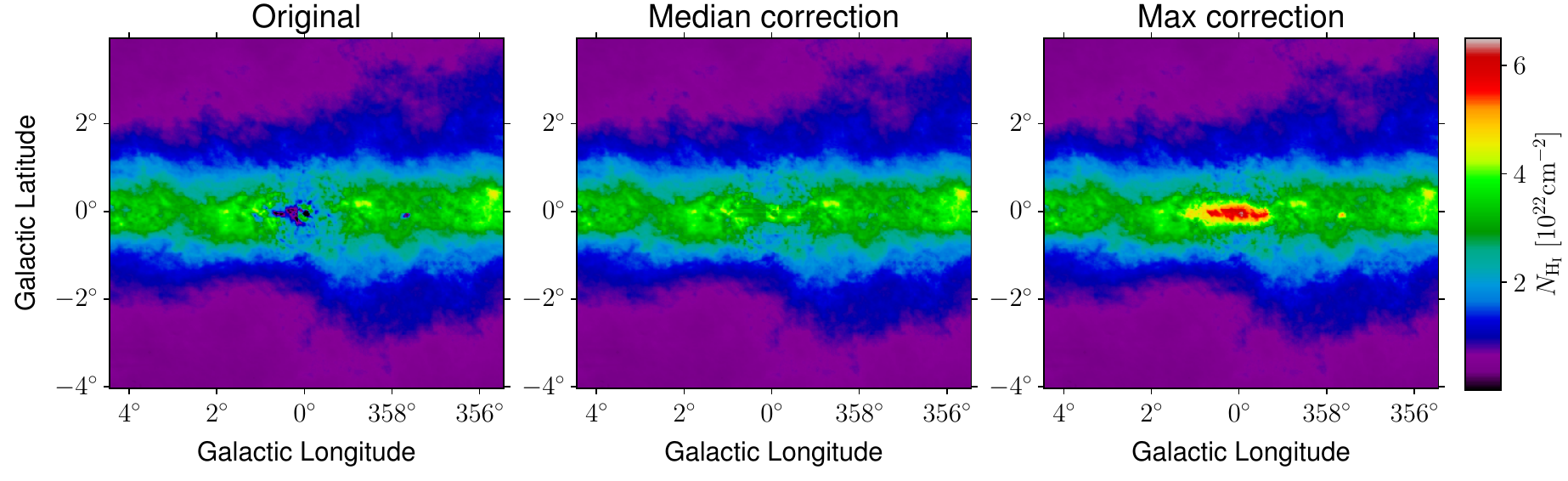}
    \caption{From left to right : \hi column density maps without correction for absorption, with median correction, and maximal correction following the method described in Sect. \ref{sec:hi_abs} and illustrated in Fig.~\ref{fig:NHImethod}.
    }
    \label{fig:NHImaps}
\end{figure}

\begin{figure}[H]
    \centering
    \includegraphics[width=.9\textwidth, trim={0.95cm 0.5cm 0.5cm 0.3cm},clip]{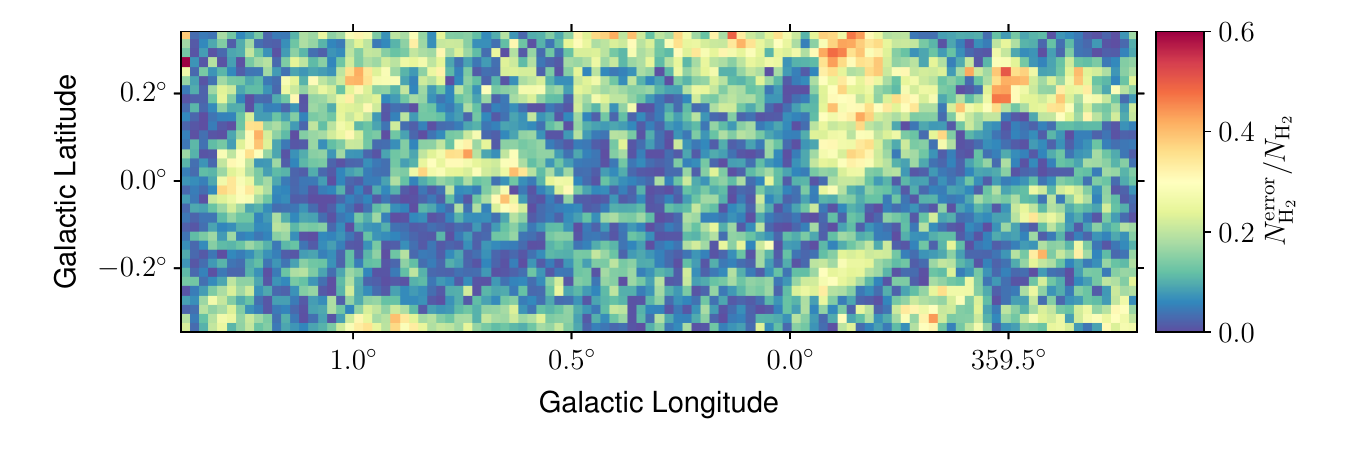}
    \caption{Relative error with respect to the average \nhd estimate (here the error is taken as the maximal deviation among the two \co isotopes considered).}
    \label{fig:err_nh2_map}
\end{figure}

\renewcommand{\arraystretch}{1.5}
\setlength{\tabcolsep}{0.3cm}

\begin{table*}[ht]
    \centering
    \caption{List of parameters used for baseline correction and line detection. In parenthesis, the corresponding number of velocity bins, $\delta \mathrm{v}$. }
    \resizebox{\textwidth}{!}{
    \begin{tabular}{c|c|c|c|c}
        \hline \hline
        \thead{Tracer} & \thead{Running mean width \\ $[\mathrm{km}\,\mathrm{s}^{-1}]$} & \thead{Gaussian smoothening width \\ $\sigma_{Gauss}$ $[\mathrm{km}\,\mathrm{s}^{-1}]$} & \thead{ Line separation \\ $\Delta v$ $[\mathrm{km}\,\mathrm{s}^{-1}]$} & \thead{ Line detection threshold \\ $\sigma_{Thr}$} \\ 
        \hline
        {\sc Hi}\xspace & - & $0.82$ ($1\delta \mathrm{v}$) & $3.3$ ($4\delta \mathrm{v}$) & $2$ \\ 
        \hline
        $^{12}\rm CO (1-0)$ & $60$ ($30\delta \mathrm{v}$) & $0.6$ ($0.3\delta \mathrm{v}$) & $4.0$ ($2\delta \mathrm{v}$) & $1$ \\
        \hline
        $^{13}\rm CO (1-0)$ & $60$ ($30\delta \mathrm{v}$) & $0.6$ ($0.3\delta \mathrm{v}$) & $4.0$ ($2\delta \mathrm{v}$) & $1$  \\
        \hline
        $^{13}\rm CO (2-1)$ & $60$ ($120\delta \mathrm{v}$) & $1.5$ ($3\delta \mathrm{v}$) & $6.0$ ($6\delta \mathrm{v}$) & $2$ \\ 
        \hline
        $\rm C^{18}O (2-1)$ & $60$ ($120\delta \mathrm{v}$) & $1.5$ ($3\delta \mathrm{v}$) & $6.0$ ($6\delta \mathrm{v}$) & $2$ \\ 
        \hline \hline
    \end{tabular}}
    \label{tab:line_parameters}
\end{table*}

\begin{figure}
\centering
     \includegraphics[width=0.85\textwidth, trim={0.3cm 1.2cm 0.64cm 0.3},clip]{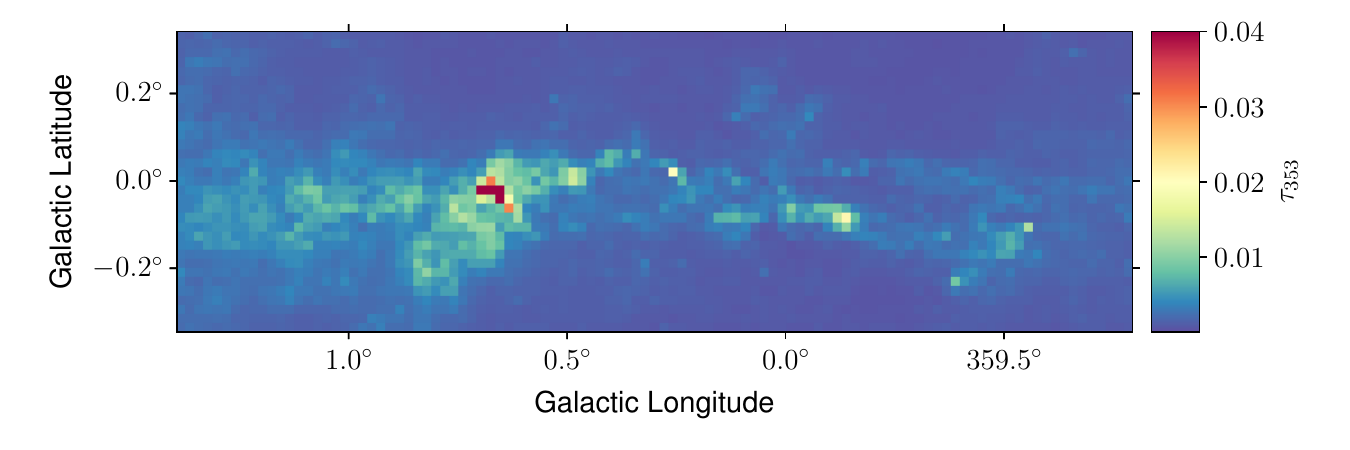}\\
     \includegraphics[width=0.85\textwidth, trim={0.55cm 1.2cm 0.64cm 0.3},clip]{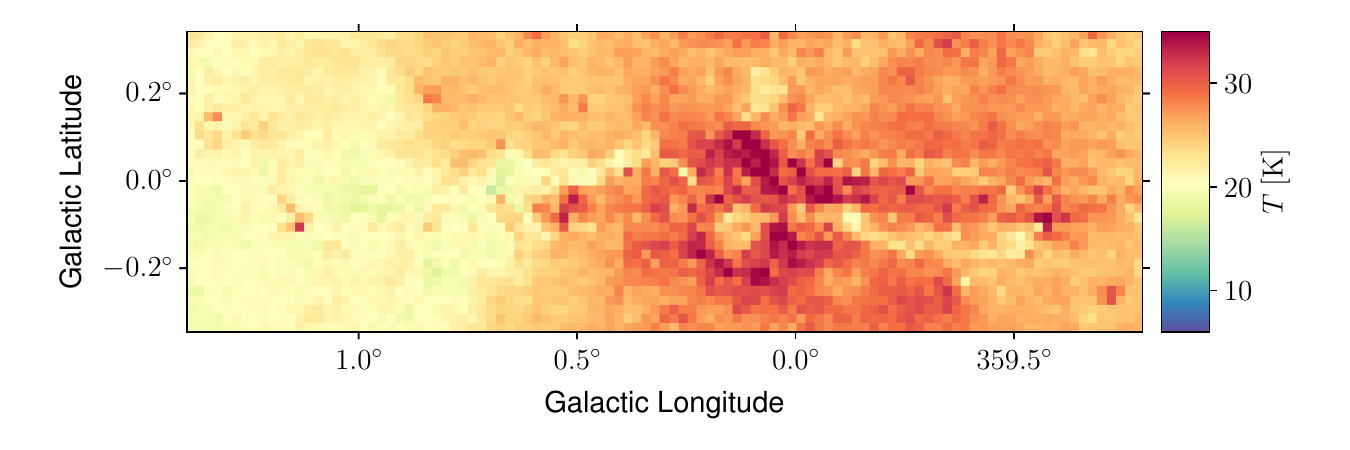}\\
     \includegraphics[width=0.85\textwidth, trim={0.55cm 0.6cm 0.64cm 0.3},clip]{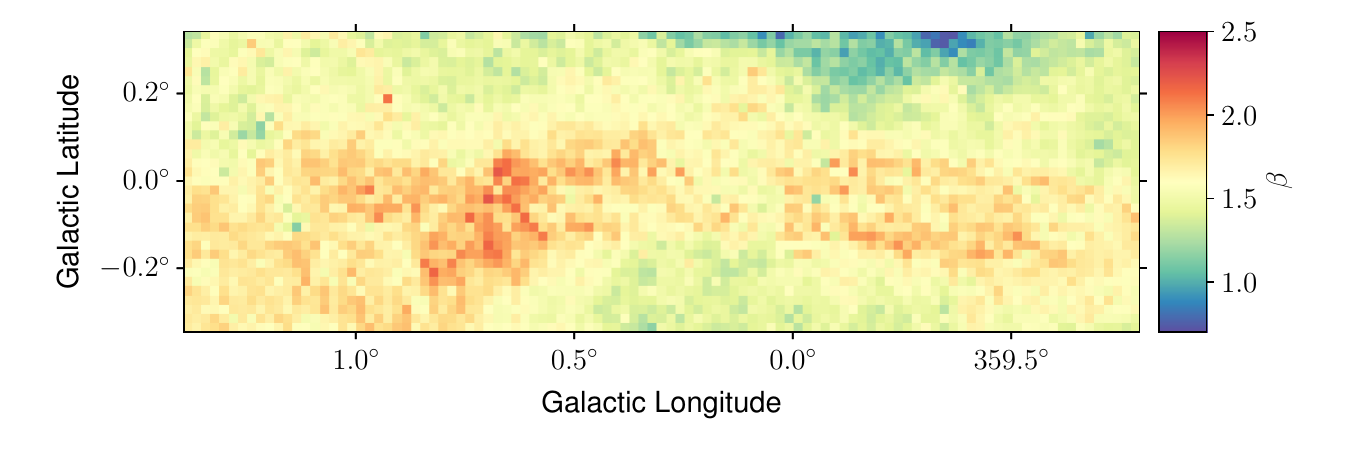}
    \caption{Parameter maps obtained by fitting the dust emission adopting a single-component model.}
    \label{fig:dust_1comp_results}
\end{figure}

\begin{figure}
\centering
     \includegraphics[width=0.85\textwidth, trim={0.3cm 1.2cm 0.64cm 0.3},clip]{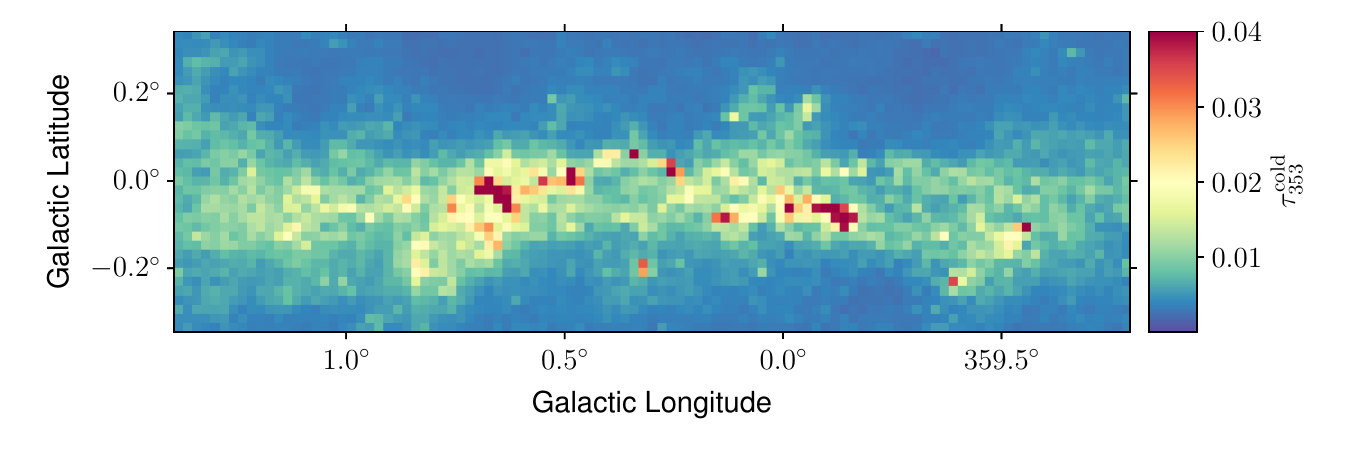}\\
      \includegraphics[width=0.85\textwidth, trim={0.3cm 1.2cm 0.64cm 0.3},clip]{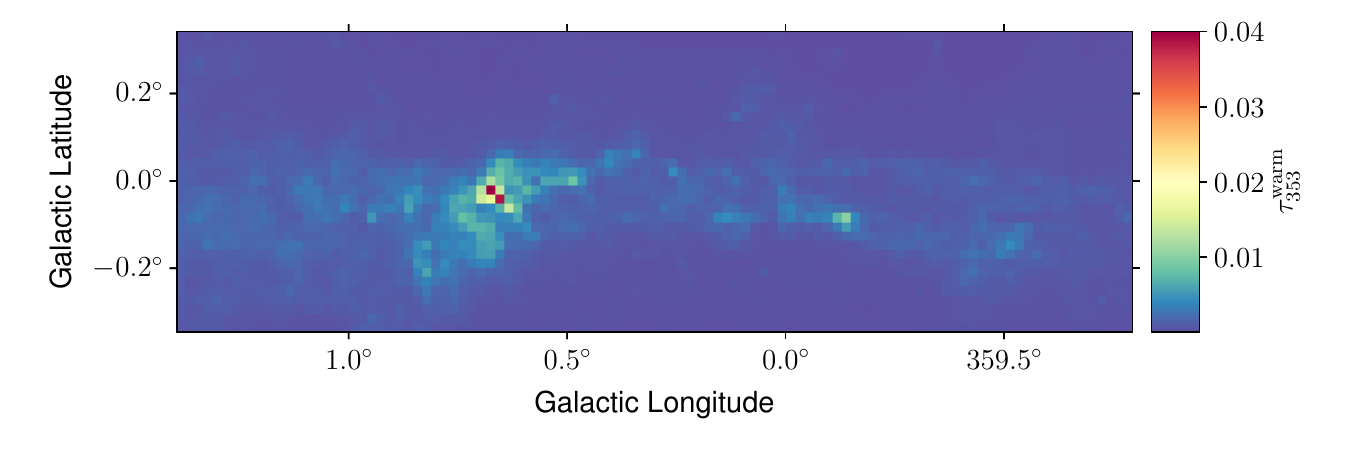}\\
     \includegraphics[width=0.85\textwidth, trim={0.55cm 1.2cm 0.64cm 0.3},clip]{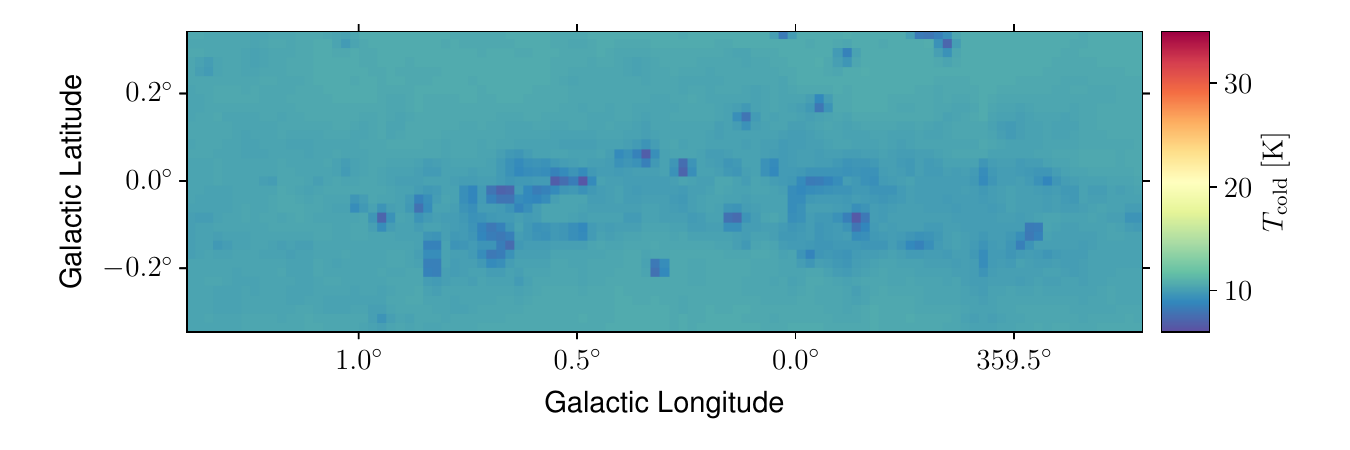}\\
      \includegraphics[width=0.85\textwidth, trim={0.55cm 0.6cm 0.64cm 0.3},clip]{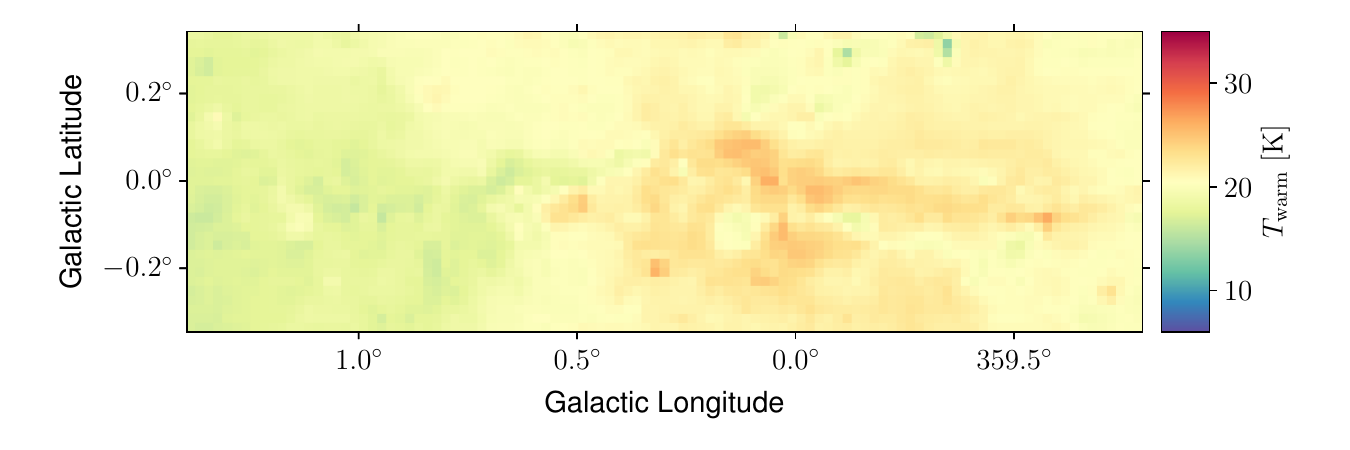}\\
     \includegraphics[width=0.85\textwidth, trim={0.55cm 0.6cm 0.64cm 0.3},clip]{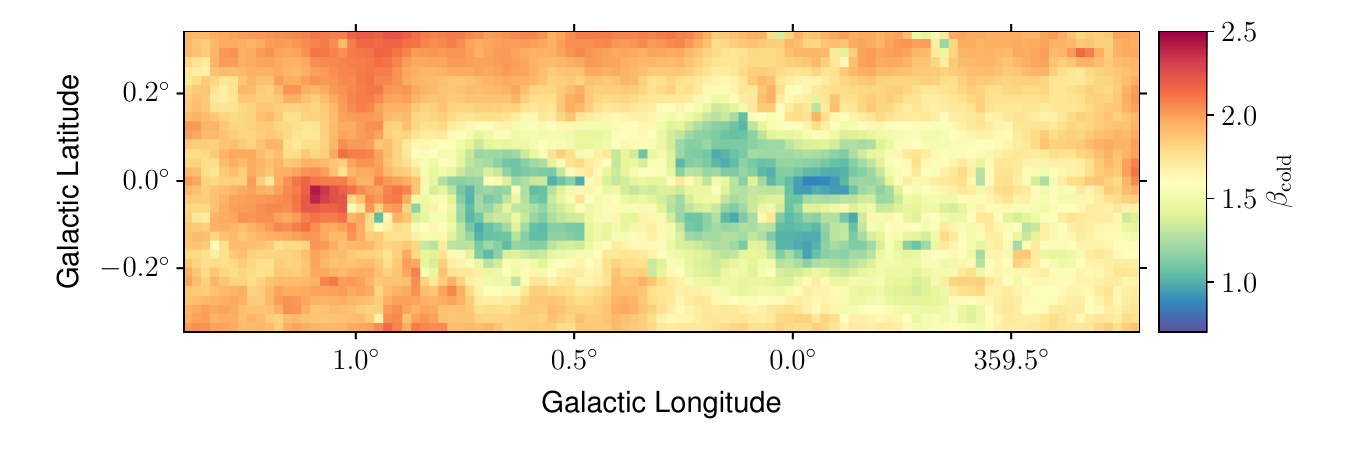}\\
    \caption{Parameter maps obtained by fitting the dust emission adopting a two-component model.}
    \label{fig:dust_2comp_results}
\end{figure}

\end{appendix}

\end{document}